\documentclass{article}
\usepackage{amsmath}
\usepackage{geometry}
\usepackage[colorlinks=true, linkcolor=blue, urlcolor=blue]{hyperref}
\usepackage{enumitem}
\usepackage{amssymb}
\usepackage{graphicx}
\usepackage{natbib}
\usepackage{amsthm}

\theoremstyle{plain}

\theoremstyle{remark}

\title{Estimation of Functional Principal Components from Sparse Functional Data}

  \author{U. MBAKA\hspace{.2cm}\\
    School of Mathematics and Statistics, University College Dublin,\\
    Belfield, Dublin, Ireland.\\
    J. CAO,\\
    Department of Statistics and Actuarial Science,\\
    Simon Fraser University, Burnaby, BC, Canada.\\
    and\\
    M. CAREY \\
    School of Mathematics and Statistics, University College Dublin,\\
    Belfield, Dublin, Ireland.}
\date{March 2026}

\begin{document}

\maketitle

\begin{abstract}
Sparse functional data arise when measurements are observed infrequently and at irregular time points for each subject, often in the presence of measurement error. These characteristics introduce additional challenges for functional principal component analysis. In this paper, we propose a new approach for extracting functional principal components from such data by combining basis expansion with maximum likelihood estimation. Orthogonality of the estimated eigenfunctions is preserved throughout the optimization using modified Gram–Schmidt orthonormalization. An information criterion is proposed to select both the optimal number of basis functions and the rank of the covariance structure. Principal component scores are subsequently estimated via conditional expectation, enabling accurate reconstruction of the underlying functional trajectories across the full domain despite sparse observations. Simulation studies demonstrate the effectiveness of the proposed method and show that it performs favorably compared with existing approaches. Its practical utility is illustrated through applications to CD4 cell count data from the Multicenter AIDS Cohort Study and somatic cell count data from Irish research dairy cattle. Supplementary materials, including technical details, additional simulation results, and the R package \texttt{mGSFPCA}, are available online.
\end{abstract}

\noindent%

\newpage

\section{Introduction}\label{sec:intro}

\begin{figure}[!htp]
    \centering
    \includegraphics[width=0.7\textwidth]{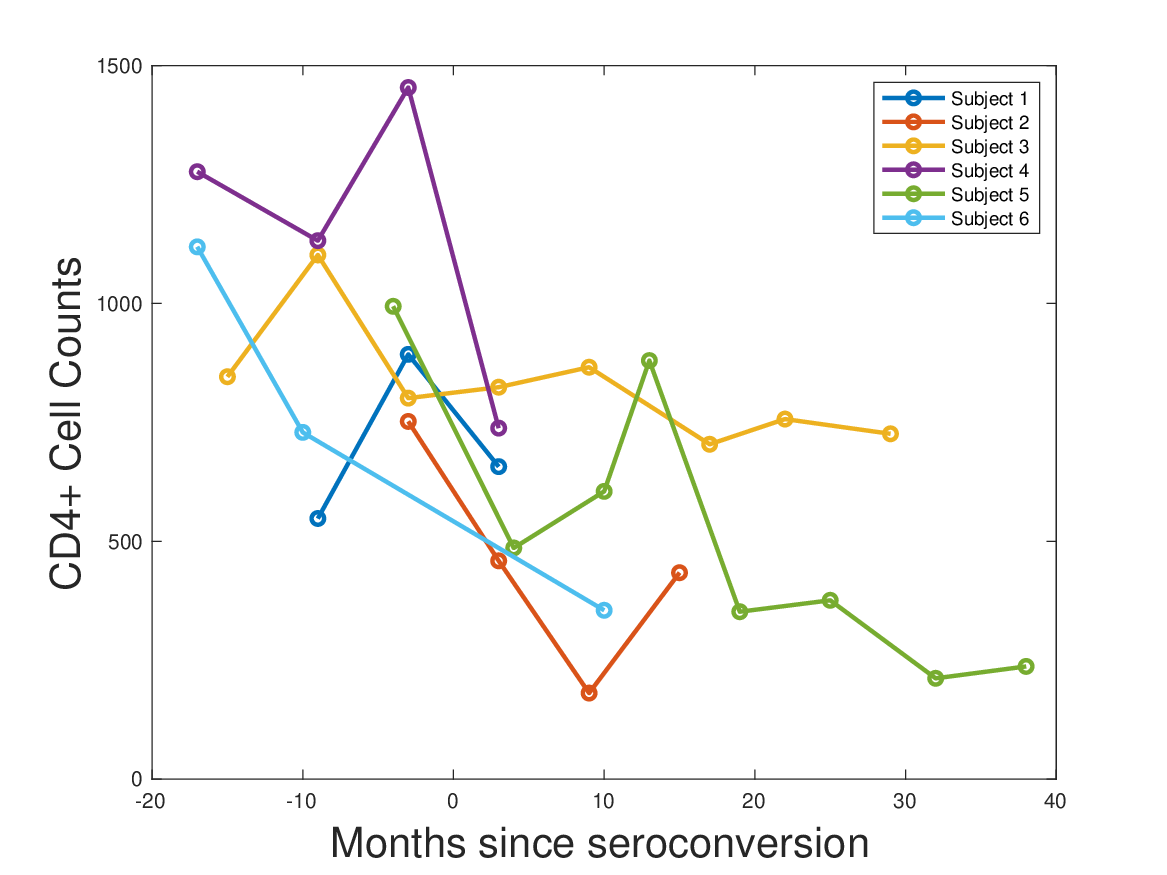}
    \caption{Spaghetti plot for sparsely recorded CD4 count data (the number of CD4 T lymphocytes in a sample of blood) for 6 subjects, each shown in a different color.}
    \label{fig:cd4}
\end{figure}

Functional Data Analysis (FDA) specializes in analyzing data that varies smoothly over a continuous domain. Most functional data are curves, \(X_i(t): t \in T\), for \(i=1, \ldots, n\), where each data unit is a function defined within the interval \(T\). One function is recorded per subject \(i\) in a random sample comprising of \(n\) individuals. Sparse Functional Data are prevalent in longitudinal studies, marked by infrequent and variable measurements per subject that are affected by measurement errors. An illustrative example of sparse functional data is shown in Figure \ref{fig:cd4}, which presents CD4 cell count data for the first six subjects in the Multicenter AIDS Cohort Study \citep{Kaslow1987}. 

As illustrated in Figure \ref{fig:cd4}, the number of measurements per subject varies significantly, ranging from three to eight observations. This is accompanied by variability in the subset of the interval \(T\) for which observations are available, as well as measurement error. The observed data for each subject, represented as \([Y_{ij}, t_{ij}]\) for \(i=1, \ldots, n,\) and \(j=1, \ldots, m_i,\) are described by:
\begin{equation}\label{eqn:data}
    Y_{ij} = X_i(t_{ij}) + e_{ij},
\end{equation}
where \(n\) is the number of subjects, \(m_i\) is the number of observations for the \(i^{th}\) subject, the \(t_{ij}\)'s are sampled from \(T\), and each error \(e_{ij}\) is assumed independent and identically distributed (i.i.d.) with a mean of zero and finite variance \(\sigma^2\).  

Functional Principal Component Analysis (FPCA) is a widely used method in FDA that identifies the principal modes of variation in functional observations. This technique enables the representation of infinite-dimensional functional data in a finite-dimensional space using an empirical basis defined by the functional principal components. FPCA has been extensively studied for scenarios where subjects are measured at the same time points as documented in \cite{Ramsay1991, Silverman1996, Ramsay2005, Ferraty2006, Hall2006, Kokoszka2017}. Furthermore, FPCA underpins advanced methods in FDA, such as functional regression \citep{Yao2005, Kraus2015, Kneip2020, Zhou2023} and functional classification and clustering \citep{James2000, Chiou2007, Stefanucci2018}. The FPCA literature is extensive and has increasingly incorporated multivariate and spatio-temporal FDA \citep{Chiou2014, Happ2018, Shi2022, Palummo2024}. This paper aims to introduce a method for extracting functional principal components from univariate sparse functional data.

We now outline the model for FPCA. Let \(\{X_i(t)\}_{i=1}^n\) represent a collection of \(n\) i.i.d square-integrable functions defined over the compact interval \(t \in T\), with a mean function \(\mu(t) = \mathbb{E}[X(t)]\) and a covariance function \(C(s,t) = \mathbb{E}[\{X(s) - \mu(s)\}\{X(t) - \mu(t)\}]\). Without loss of generality, let $T=[0,1]$. As \(X(t)\) is an \(L^2\)-stochastic process, the covariance function \(C(s,t)\) can be expressed through an orthogonal expansion according to Mercer's Theorem \citep{Mercer1909}. This expansion involves a sequence of eigenfunctions \(\{\phi_k(t)\}_{k=1}^\infty\) and non-increasing eigenvalues \(\{\lambda_k\}_{k=1}^\infty\), such that 
\begin{equation}\label{Mercer}
C(s,t) = \sum_{k = 1}^\infty \lambda_k \phi_k(s) \phi_k(t) \quad \textrm{for} \quad s,t \in T.
\end{equation}
The functions \(\phi_k\) form an orthonormal basis, satisfying \(\int \phi_k(t)^2 dt = 1\) and \(\int \phi_k(t) \phi_l(t) dt = 0\) for \(l \neq k\), see \cite{Hsing2015} for example. Consequently, the \(i^{th}\) random function is characterized by a Karhunen–Loève representation \citep{loeve1945fonctions, karhunen1947under} given by:
\begin{equation}\label{true_func}
    X_i(t) = \mu(t) + \sum_{k = 1}^\infty \xi_{ik} \phi_k(t), 
\end{equation}
where the random variables, called the scores, $\xi_{ik}$ are uncorrelated with zero mean and variance $\lambda_k$. 

Various methodologies have been developed to estimate the leading \( p \) eigenfunctions \(\{\phi_k(t)\}_{k=1}^p\), their corresponding eigenvalues \(\{\lambda_k\}_{k=1}^p\), and the error variance \(\sigma^2\) from sparse functional data. As proposed by \cite{Shi1996} and \cite{James2000}, a reduced rank mixed-effects approach models each trajectory using B-splines with random coefficients. A prevalent indirect method involves a two-stage process: initially, data from all subjects are pooled, and a smoothing technique is applied to estimate \(\mu(t)\) and \(C(s,t)\). Subsequently, the eigenfunctions and eigenvalues are estimated through a spectral decomposition of the estimated covariance \(\hat{C}(s,t)\). In this context, \cite{Staniswalis1998, Yao2003, Yao2005} advocated for the use of a local polynomial regression smoother, while \cite{Xiao2018} employed bivariate penalized spline smoothing to obtain \(\hat{C}(s,t)\).

\cite{Peng2009} introduced a direct approach that approximates the eigenfunctions in (\ref{Mercer}) using a known finite basis of smooth functions. By maximizing the restricted log-likelihood on a Stiefel manifold, this method also ensures that the estimated basis function expansion of the eigenfunctions forms an orthonormal basis. A penalized version employing conjugate gradient optimization was implemented by \cite{He2022}. Another direct approach, proposed by \cite{Cai2010}, assumes that the sample path of \(X\) belongs to a reproducing kernel Hilbert space (RKHS), where \(X(t) = \sum_{k \geq 1} \xi_k \phi_{k}(t)\), with \(\xi_k = \langle X, \phi_{k} \rangle_{L_2}\) representing the Fourier coefficient of \(X\) with respect to \(\{\phi_{k}(t): k \geq 1\}\), and \(\phi_{k}(t)\) being the eigenfunctions of the reproducing kernel's eigenvalue decomposition. Furthermore, \cite{Nie2022} proposed an approach for recovering the underlying individual trajectories as well as the major variation patterns denoted by the eigenfunctions \(\{\phi_k(t)\}_{k=1}^p\) using a finite approximation of (\ref{true_func}). This method extends the approach of \cite{Huang2008} to sparse or longitudinal data. 

Separately, \cite{Kneip2020} and \cite{Kraus2015} address the estimation of principal component scores in (\ref{true_func}) for partially observed functional data under the assumption that the eigenfunctions and eigenvalues are available from standard estimation procedures (e.g. \cite{Yao2005}).

When the functions are observed as fragments or snippets, the methods proposed by \cite{Delaigle2016, Descary2019} and \cite{Lin2022} can be used to estimate the covariance function. Fragmented functional data involves large portions of the domain that are never jointly observed for any subject, leading to methodological challenges.

The contributions of this article are as follows. Following \cite{Peng2009}, we adopt a reduced-rank model for the covariance function, approximating Mercer's infinite expansion in (\ref{Mercer}) using a truncated series. This method efficiently captures the dominant modes of variation in the sparse functional data with minimal bias, even with a limited number of principal components.
Despite its theoretical appeal, optimization on the Stiefel manifold is widely recognized as numerically challenging \citep{Absil2008, Li2020, Kume2021, NachuanXiao2024}. In particular, the Newton–Raphson method adopted by \cite{Peng2009}, based on the framework of \cite{Edelman1998}, can be sensitive to initialization because of the manifold’s non-convex geometry and may exhibit slow convergence in sparse data regimes where the likelihood surface is poorly conditioned \citep{Absil2008, Peng2009, JonathanWSiegel2021}. While \cite{He2022} also sought to overcome these challenges by proposing a conjugate gradient algorithm over the product manifold, their approach remains within the complex geometric optimization framework we aim to avoid. In contrast, our approach enforces orthonormality in the basis expansion of the eigenfunctions by incorporating modified Gram–Schmidt (MGS) orthonormalization iteratively during estimation. This strategy circumvents direct optimization of the coefficients over the Stiefel manifold, thereby simplifying the estimation procedure and improving numerical stability and computational efficiency. Standard optimization techniques can then be employed, with the orthonormality of the estimated eigenfunctions preserved through the MGS orthonormalization. We use a quasi-Newton method to minimize the negative log-likelihood, yielding estimates of the basis coefficients for the eigenfunctions, the eigenvalues, and the error variance. To determine the optimal number of basis functions and principal components, we use an AIC-type selection criterion. Finally, by employing conditional expectation, we estimate the principal scores and successfully recover the underlying trajectory across the entire domain from the sparse functional data. The code for the proposed method is publicly available at \href{https://github.com/uchembaka/mGSFPCA}{Github}

The paper is organized as follows: Section \ref{method} describes our proposed method for estimating the principal eigenfunctions, corresponding eigenvalues, and error variance from sparsely observed functional data. Section \ref{sims} presents simulation studies comparing our method to the Restricted Maximum Likelihood Estimator \citep{Peng2009}, the conjugate gradient approach \citep{He2022}, the spectrally regularized covariance estimator in a Reproducing Kernel Hilbert Space \citep{Wong2019}, the local polynomial regression smoother in Principal Component Analysis through Conditional Expectation \citep{Yao2005}, Sparse Orthonormal Approximation  \citep{Nie2022}, and Fast Covariance Estimation  \citep{Xiao2018}. In Section \ref{apps}, we apply our method to the CD4 cell count data from the Multicenter AIDS Cohort Study \citep{Kaslow1987} and somatic cell count data of dairy cattle across various research farms
in Ireland. Section \ref{sec:conc} concludes with a discussion and outlines directions for future research.

\section{Methodology}\label{method}

Assume a reduced-rank model for the covariance function, approximating the infinite Mercer's expansion in (\ref{Mercer}) with a truncated version 
\[C_p(s,t) = \sum_{k = 1}^p \lambda_k \phi_k(s) \phi_k(t)\]
Due to the rapid decay of eigenvalues, the bias from truncation is minimal even for small \(p\) \citep{James2001, Peng2009}. Thus, the random function for the \(i^{th}\) individual can be approximated by:
\begin{equation}\label{Aprox_X}
    \hat{X}_i(t)=\hat{\mu}(t) + \sum_{k = 1}^p \hat{\xi}_{ik} \hat{\phi}_k(t),
\end{equation}
where \(\hat{\xi}_{ik}\) is the \(k^{th}\) estimated principal component score, and \(\hat{\phi}_k(t)\) is the \(k^{th}\) estimated eigenfunction. The mean function \(\mu(t)\) is estimated by applying cubic spline interpolation to the pooled observations from all individuals, a standard approach in FDA \citep{Ramsay2005}. The principal component scores \(\{\xi_{ik}\}_{k=1}^p\) are estimated through conditional expectation. In particular, assuming the observations $\{Y_{ij}\}_{j=1}^{m_i}$ are normally distributed, the scores for the $i^{th}$ subject are obtained from: 
\begin{equation}\label{CE} 
\mathbb{E}[\boldsymbol{\xi}_i | \mathbf{Y}_i] = \boldsymbol{\Lambda}\boldsymbol{\phi}_i^\top \boldsymbol{\Sigma}_i^{-1} (\mathbf{Y}_i - \boldsymbol{\mu}_i), 
\end{equation} 
where $\boldsymbol{\Lambda}$ is a $p \times p$ diagonal matrix containing the eigenvalues $\{\lambda_k\}_{k=1}^{p}$, and $\boldsymbol{\phi}_i$ is an $m_i \times p$ matrix of eigenfunctions $\phi_k(t)$ evaluated at the observed time points $\mathbf{t}_i$. The vector $\mathbf{Y}_i$ represents the $m_i$ observations for the subject, while $\boldsymbol{\mu}_i$ is the corresponding vector obtained by evaluating the mean function $\mu(t)$ at those same time points. The $m_i \times m_i$ matrix $\boldsymbol{\Sigma}_i$ represents the reduced rank covariance of the observations, with the $(j,l)^{th}$ element defined as $\Sigma_{i,j,l}=C(t_{i,j},t_{i,l})+\mathbf{I}_{j=l}\sigma^2$, where $\sigma^2$ is the error variance \citep{Yao2005}. The asymptotic properties of this estimator were established in \cite{Yao2005}.

The $k^{th}$ eigenfunction of (\ref{Aprox_X}), $\phi_k(t)$, is modelled using a basis expansion: $\phi_k(t) = \sum_{l = 1}^Q \tilde{c}_{lk}\tilde{B}_{l}(t)$, where $\{\tilde{B}_{l}(\cdot)\}_{l=1}^Q$ (with $Q \geq p$) are orthonormal basis functions and $\boldsymbol{\tilde{C}} = ((\tilde{c}_{kl}))$ is a $Q \times p$ coefficient matrix satisfying $\boldsymbol{\tilde{C}}^\top\boldsymbol{\tilde{C}} = \boldsymbol{I}_p$. \cite{Peng2009} and \cite{He2022}, directly estimate $\boldsymbol{\tilde{C}}$ under this orthogonality constraint, which requires optimization on the Stiefel manifold, a procedure that can be numerically unstable. An alternative reparameterization approach is proposed herein.

An unconstrained basis expansion is initially constructed for the $k^{th}$ eigenfunction, $$u_k(\boldsymbol{\tau}) = \sum_{l = 1}^Q c_{kl} B_{kl}(\boldsymbol{\tau}) = \boldsymbol{B}\boldsymbol{c_k}.$$ Here, $\boldsymbol{c_k} \in \mathbb{R}^{Q}$ is a coefficient vector, and $\boldsymbol{B} \in \mathbb{R}^{M\times Q}$ is a matrix of basis functions (e.g., B-splines or Fourier basis, see \cite{Ramsay2005}) evaluated on a dense grid of $M$ quadrature points
$\boldsymbol{\tau} = \{\tau_1,\ldots,\tau_M\} \subset [0,1]$. At each iteration of the optimization procedure, the vector-valued functions $\textbf{U} = [\boldsymbol{u_1},\ldots,\boldsymbol{u_p}]$ are orthonormalized directly. Specifically, a modified Gram–Schmidt (MGS) process is applied to $\textbf{U}$ to produce an orthonormal set of eigenfunctions $\boldsymbol{\Phi} = [\phi_1(\boldsymbol{\tau}), \ldots, \phi_p(\boldsymbol{\tau})] \in \mathbb{R}^{M \times p}$. The MGS procedure enforces orthonormality on the discrete grid $\boldsymbol{\tau},$ which corresponds to orthogonality under a discrete inner product. To ensure consistency with the continuous $L^2[0,1]$ inner product the MGS recursion is formulated using a quadrature based discretized inner product,
$
\langle f, g \rangle \approx \sum_{j=1}^{M} w_j f(\tau_j)g(\tau_j) \equiv \langle f, g \rangle_w,
$
where \(\{w_j\}_{j=1}^{M}\) denotes the quadrature weights. In practice, these weights are computed using the trapezoidal rule, which yields an approximation error of the order $\mathcal{O}((M-1)^{-2})$ for sufficiently smooth functions as the grid $\boldsymbol{\tau}$ becomes dense. The corresponding norm is \(\|f\|_w = \sqrt{\langle f, f \rangle_w}\). The orthonormal matrix of the MGS $\mathcal{M}_W : \mathbb{R}^{M \times p} \to \mathbb{R}^{M \times p}$ is denoted by $\boldsymbol{\Phi} = \mathcal{M}_W(\textbf{U})$ and defined by
\begin{equation}\label{MGS_Ap}
\phi_k(\boldsymbol{\tau}) =\frac{
u_k(\boldsymbol{\tau}) - \sum_{\nu=1}^{k-1} \langle u_k, \phi_\nu \rangle_w \, \phi_\nu(\boldsymbol{\tau})}{
\left\| u_k(\boldsymbol{\tau}) - \sum_{\nu=1}^{k-1} \langle u_k, \phi_\nu \rangle_w \, \phi_\nu(\boldsymbol{\tau}) \right\|_w}.
\end{equation}
The resulting eigenfunctions $\phi_k$ form an orthonormal basis, leading to the $M \times M$ reduced-rank covariance matrix:
\begin{equation}\label{Cov_Est}
        \boldsymbol{\Sigma} = \underbrace{\mathcal{M}_W({\textbf{B}\textbf{C}})}_{\boldsymbol{\Phi}} \boldsymbol{\Lambda} \underbrace{\mathcal{M}_W({\textbf{B}\textbf{C}})^{\top}}_{\boldsymbol{\Phi}^{T}} + \sigma^2 \boldsymbol{I}_{M},
\end{equation}
where $\sigma^2$ is the error variance, $\boldsymbol{\Lambda} = \text{diag}(\lambda_1, \ldots, \lambda_p)$ is the $p \times p$ matrix of eigenvalues and $\boldsymbol{I}_{M}$ is the $M \times M$ identity matrix. The formulation in (\ref{Cov_Est}) is asymptotically equivalent to the model in \cite{Peng2009} and \cite{He2022} see Appendix A.1 for details. 

A maximum likelihood approach is used to estimate the basis expansion coefficients \(\boldsymbol{C} = [\boldsymbol{c}_{1}, \ldots, \boldsymbol{c}_{p}] \in \mathbb{R}^{Q \times p}\), the error variance \(\sigma^2\), and the eigenvalues \(\{\lambda_k\}_{k=1}^{p}\). At each iteration of the optimization, the current coefficient estimates are used to construct the functions \(u_k(\boldsymbol{\tau})\), which are then orthonormalized via the MGS procedure described in (\ref{MGS_Ap}) to obtain the vector-valued functions \(\phi_k(\boldsymbol{\tau})\). For optimization purposes, the transformed parameters \(\gamma = \log \sigma^2\), ensuring positivity of the error variance and \(\eta_k = \log \lambda_k\), ensuring positivity of the eigenvalues, are introduced. The average negative log-likelihood of the observed data, conditional on \(\{(m_{i},\{t_{i,j}\}_{j=1}^{m_{i}})\}_{i=1}^n\), is given by:
\begin{equation}\label{eqn:LLK}
    \mathcal{L} = -\log L(\textbf{C}, \boldsymbol{\Lambda}, \sigma^2) = \frac{1}{n} \sum_{i = 1}^n \text{tr} [ \boldsymbol{\Sigma}_{i}^{-1} (\textbf{Y}_i-\boldsymbol{\mu}_i)(\textbf{Y}_i-\boldsymbol{\mu}_i)^{\top}] + \frac{1}{n} \sum_{i=1}^n \log|\boldsymbol{\Sigma}_{i}|.
\end{equation}
where \(\boldsymbol{\Sigma}_{i}\) is the covariance matrix obtained by evaluating (\ref{Cov_Est}) at the observation times \(\textbf{t}_i\) for the \(i^{th}\) individual. The minimization of \eqref{eqn:LLK} lead to the estimators
\begin{equation}\label{eqn:mGSFPCA_estimator}
    (\hat{\textbf{C}}, \hat{\boldsymbol{\Lambda}}, \hat{\sigma}^2) = \arg\min_{C \in \mathbb{R}^{Q \times p},\; \Lambda \in \mathcal{D}_+,\; \sigma^2 > 0} \mathcal{L}\bigl(\textbf{C},\boldsymbol{\Lambda},\sigma^2\bigr),
\end{equation}
where $\mathcal{D}_+ = \{\mathrm{diag}(\lambda_1,\ldots,\lambda_p): \lambda_k > 0\}.$ The minimization in (\ref{eqn:mGSFPCA_estimator}) involves standard numerical optimization techniques such as the BFGS Quasi-Newton method combined with a cubic line search \citep{Fletcher2000}. The required score functions admit closed-form expressions, which enable efficient implementation and are provided in Appendix~A.2. Convergence of the BFGS Quasi-Newton method combined with a cubic line search in this context is established in Appendix~A.3. Practical aspects of the optimization, including parameter initialization, convergence criteria, and numerical stabilization, follow standard strategies and are summarized in Appendix~A.4. The resulting estimate of the eigenfunctions is $\hat{\boldsymbol{{\Phi}}}=\mathcal{M}_W(\textbf{B}\hat{\textbf{C}}).$

Once the eigenfunctions have been estimated on the evaluation grid using the parameter estimates obtained by minimizing~(\ref{eqn:LLK}), a continuous functional representation can be obtained by drawing on classical results from spline approximation theory \citep{DeBoor1978}. Using the symmetric inverse square root $\mathbf{G}_B^{-1/2}$ of the $Q \times Q$ Gram matrix $\mathbf{G}_B = \int_0^1 \boldsymbol{B}(t)\boldsymbol{B}(t)^\top\,dt$, we define the orthonormal B-spline basis evaluation at any $t \in T$ as $\tilde{\boldsymbol{B}}(t) = \mathbf{G}_B^{-1/2}\boldsymbol{B}(t)$, where $\boldsymbol{B}(t) = [B_1(t),\dots,B_Q(t)]^\top$ is the length-$Q$ vector of standard B-spline basis evaluations at $t$. The coefficient matrix $\tilde{\mathbf{C}}$ is then obtained by 
\begin{equation}\label{Ctilde}
\tilde{\textbf{C}} = \int_0^1 \tilde{\mathbf{B}}(t) \boldsymbol{\Phi}(t) \textrm{d}t \approx \tilde{\mathbf{B}}^\top \mathbf W \boldsymbol{\Phi},
\end{equation}
where $\tilde{\mathbf{B}}$ is the $M \times Q$ matrix of orthonormal basis evaluations at the quadrature points and $\mathbf W$ is the $M\times M$ diagonal matrix of trapezoidal weights. Then for any $t \in T$, the $1\times p$ vector $\boldsymbol{\Phi}(t)$ can then be evaluated as
\begin{equation}\label{eqn:phi_t}
\boldsymbol{\Phi}(t) = \tilde{\mathbf{B}}(t)^\top \tilde{\mathbf{C}}.
\end{equation}
\subsection{Parameter Selection for  \texorpdfstring{$Q$}{Q} and \texorpdfstring{$p$}{p}}

To select the number of basis functions \(Q\) and the number of principal components \(p\), a model selection criterion is required. Although cross-validation is a natural candidate, its computational cost is often prohibitive, motivating the use of efficient approximations such as information criteria.

Standard information criteria face both conceptual and practical challenges in our setting. AIC \citep{akaike1974new} with a parameter count of \(Qp + p + 1\) assumes independence among parameter estimates, which is violated by the sequential dependence induced by the MGS procedure. TIC \citep{takeuchi1976distribution} corrects for misspecification by replacing the fixed penalty with $\textrm{tr}(\mathbf{I}^{-1}\mathbf{J}),$ where \(\mathbf{I}\) is the empirical Fisher information operator and \(\mathbf{J}\) is the empirical score covariance \citep{burnham2002model}. However, as stated in \cite{Peng2009} it is unclear an approximation to this penalty is appropriate in the current context because its implicit assumptions about asymptotic behavior of the empirical Fisher information operator have not been verified.

Given these challenges, we instead develop a heuristic penalty informed by the structure of the estimation procedure. The reduced-rank covariance model takes the form $\boldsymbol{\Sigma} = \sum_{k=1}^{p} \lambda_k \boldsymbol{\phi}_k \boldsymbol{\phi}_k^\top + \sigma^2 \mathbf{I}$, where each eigenfunction \(\boldsymbol{\phi}_k = \mathcal{M}_W(\mathbf{B}\mathbf{c}_k)\) is obtained by applying the weighted MGS procedure to the basis expansion \(\mathbf{B}\mathbf{c}_k\). As shown in \eqref{MGS_Ap}, a perturbation in a coefficient vector \(\mathbf{c}_j\) affects not only \(\boldsymbol{\phi}_j\) but all subsequent eigenfunctions \(\boldsymbol{\phi}_k\) with \(k > j\). Thus, the coefficient vectors are not independent. Moreover, each of the \(Q\) components of \(\mathbf{c}_j\) may influence up to \(p-j+1\) eigenfunctions. Summing this influence over \(j\) gives \(\sum_{j=1}^p (p-j+1) = p(p+1)/2 = O(p^2)\). Consequently, the total number of coefficient-eigenfunction interactions grows on the order of \(O(Qp^2) \).

We propose an approximate model complexity of \(Qp^2 + p+ 1\), where the additional terms account for the eigenvalues and noise variance. 
This leads to an AIC given by
\begin{equation}\label{eqn:GCV}
\mathrm{AIC} = n\mathcal{L}(\hat{\mathbf{C}},\hat{\lambda}_1,\dots,\hat{\lambda}_p,\hat{\sigma}^2) + (Qp^2 + p + 1),
\end{equation}
where \(\mathcal{L}\) is the average negative log-likelihood. We emphasize that this choice is a heuristic approximation of model complexity motivated by the model structure, not an exact derivation; the precise effective degrees of freedom have not been derived for this constrained and non-linear setting. Nevertheless, the proposed penalty is supported by the simulation studies presented in Section \ref{sec:model_sel}, where it consistently selects parsimonious models and achieves performance comparable to cross-validation.

\subsection{A Confidence Interval for \texorpdfstring{\(X_i(t)\)}{X} }

As shown in (\ref{Aprox_X}), $\hat{X}_i(\textbf{t}_{i})$ is approximated using the first $p$ leading functional principal components. Following \cite{Yao2005}, the \((1-\alpha)\) asymptotic point-wise confidence interval for \(X_i(\textbf{t}_{i})\) is given by:
\begin{equation}\label{eqn:CI}
    \hat{X}_i(\textbf{t}_{i}) \pm \gamma \sqrt{ \boldsymbol{\phi}(\textbf{t}_{i})^\top [ \hat{\boldsymbol{\Lambda}} - \hat{\textbf{H}} \hat{\boldsymbol{\Sigma}}_i^{-1} \hat{\textbf{H}}^\top ] \boldsymbol{\phi}(\textbf{t}_{i}) },
\end{equation}
where \(\gamma\) is the \(1-\frac{\alpha}{2}\) quantile from the standard Gaussian cumulative distribution function, and \(\hat{\textbf{H}}\) is a $p \times n$ matrix containing \((\hat{\lambda}_1 \hat{\phi}_{i1}, \ldots, \hat{\lambda}_p \hat{\phi}_{ip})^\top\). The consistency and asymptotic properties of (\ref{eqn:CI}) are in \cite{Yao2005} and \cite{Hall2006}.

\section{Simulation Study}\label{sims}

\begin{figure}[htp]
    \centering
    \includegraphics[width=0.95\textwidth]{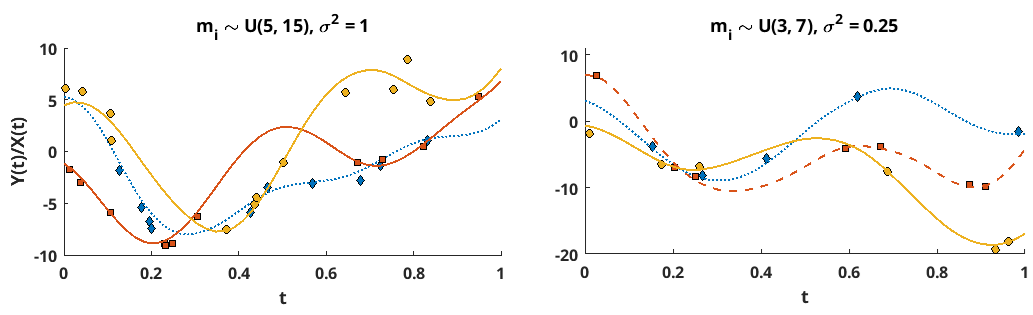}
    \caption{Simulated functional data with observed points \(Y(t)\) (dots) and true curves \(X(t)\) (lines). Left: The Matérn covariance with \(m_i \sim U_{d}(5,15)\) and \(\sigma^2 = 1\). Right: The Egg Crate covariance with \(m_i \sim U_{d}(3,7)\) and \(\sigma^2 = 0.25\).}
    \label{fig:sim}
\end{figure}

A simulation study was conducted under two distinct covariance structures: the Mat\'{e}rn Covariance \(\Sigma(s,t) = \sigma^2 \frac{2^{1-\nu}}{\Gamma(\nu)}\left(\sqrt{2\nu} \frac{|s-t|}{\rho}\right)^{\nu} K_{\nu} \left(\sqrt{2\nu} \frac{|s-t|}{\rho}\right)\) and a finite basis expansion \(\Sigma(s,t) = \sum^3_{k=1}\lambda_k \phi_k(s) \phi_k(t) \), where \(\phi_k\)'s are the eigenfunctions and \(\lambda_k\)'s are the eigenvalues. For each covariance structure, the three parameter settings in Table \ref{Sim:Exp} were examined resulting in six simulation scenarios. Across all scenarios, the observation times $t_{i,j}$ were independently sampled from $U(0,1)$.  The number of observations per curve, $m_i$, was drawn from a discrete uniform distribution $U_{d}(l,u)$, where $l$ and $u$ denote the lower and upper bounds, respectively. The error terms $e_{i,j}$ were independently sampled from $N(0,\sigma^2)$, 
yielding observations of the form
$
Y_{ij} = X_i(t_{ij}) + e_{i,j}.
$ 
For each scenario, 100 replicates were generated. Representative examples of simulated data, together with the corresponding true curves under the two covariance structures, are shown in Figure~\ref{fig:sim}.
\begin{table}[ht]
\centering
\begin{tabular}{cc}
\hline
\textbf{Setting} & \textbf{Parameters} \\
\hline
Setting 1 & $n = 50$, $l=5$, $u=15$, $\sigma^2 = 1.$ \\

Setting 2 & $n = 100$, $l=5$, $u=15$, $\sigma^2 = 1.$\\

Setting 3 & $n = 500$, $l=3$ $u=7$, $\sigma^2 = 0.25.$\\
\hline
\end{tabular}
\caption{Overview of parameter settings for the simulation scenarios.}
\label{Sim:Exp}
\end{table}

For the Mat\'{e}rn covariance scenarios, we used the mean function $\mu(t) = 5\cos(4t^3 + 6t^2 - 12t)$ and Mat\'{e}rn covariance parameters $\sigma = 1$, $\rho = 0.1$, and $\nu = 4$ \citep{Ramsay2017}. For the finite basis expansion covariance (henceforth Egg Crate), we used the mean function $\mu(t) = 5\sin(2\pi t)$, with eigenfunctions $\phi_1(t) = \sqrt{2}\sin(2 \pi t)$, $\phi_2(t) = \sqrt{2}\cos(4 \pi t)$, $\phi_3(t) = \sqrt{2}\sin(4 \pi t)$, and eigenvalues $\lambda_1=1$, $\lambda_2=0.5$, and $\lambda_3=0.25$ \citep{Xiao2018}.
An illustration of the population means and covariances for the stochastic processes described above and used to generate the data is presented in Appendix A.5.1.

Using a dense grid \(\mathbf{g}\) of 50 equally spaced evaluation points spanning the interval $[0, 1]$, the proposed method was compared with several existing approaches. These include direct methods: Restricted Maximum Likelihood (\texttt{ReMLE}) \citep{Peng2009}, the conjugate gradient approach (\texttt{CG}) \citep{He2024}, the spectrally regularized covariance estimator (\texttt{RKHS}) \citep{Wong2019}, and Sparse Orthonormal Approximation (\texttt{SOAP}) \citep{Nie2022}. In addition, we compared against indirect methods: Principal Component Analysis through Conditional Expectation (\texttt{loc}) \citep{Yao2005} and Fast Covariance Estimation (\texttt{FACE}) both the single-stage \texttt{FACE1} and double-stage \texttt{FACE2} variants \citep{Xiao2018}. For our method, we employed B-spline basis functions for the Mat\'{e}rn covariance scenarios and Fourier basis functions for the Egg Crate scenarios. In both cases, we selected the number of basis functions, \((Q \in \{5, \cdots, 11\})\), and the number of principal components, 
\((p \in \{2, \cdots, 6\})\), by performing a grid search over all valid $(Q,p)$ combinations and choosing the pair that minimized the information criterion in (\ref{eqn:GCV}). The parameter specifications for the competing methods are provided in Appendix A.5.2.
For clarity and to ensure consistent terminology in the presentation of our simulation results, we refer to the proposed method as \texttt{mGSFPCA}.

\subsection{The Eigenfunctions and EigenValues}
Accuracy of the estimated eigenfunctions and eigenvalues was assessed using two metrics: the Root Mean Squared Error (RMSE) for the eigenfunctions, defined as  \[\text{RMSE}_{\phi_k} = \min\left[\sqrt{\frac{1}{50} \sum_{j=1}^{50} (\hat{\phi}_k(g_j) - \phi_k(g_j))^2}, \sqrt{\frac{1}{50}\sum_{j=1}^{50} (\hat{\phi}_k(g_j) + \phi_k(g_j))^2}\right],\] and the Squared Error (SE) for the eigenvalues, given by \( \text{SE}_{\lambda_k} = (\hat{\lambda}_k - \lambda_k)^2 \). 

Table \ref{table:rmse_phi} presents the median and interquartile range (IQR) of \(100 \times \text{RMSE}_{\phi_k}\) for the top three principal eigenfunctions (\(\phi_1, \ \phi_2,\) and \(\phi_3\)) across the six simulated scenarios. To ensure a fair comparison, we excluded simulation replicates where the \texttt{ReMLE} or \texttt{CG} methods failed to converge. For the Mat\'{e}rn covariance scenarios, 96, 86, and 81 out of 100 replicates converged for \texttt{ReMLE} in Settings 1, 2, and 3, respectively. For the \texttt{CG} method, 19 out of 300 total replicates across the three Mat\'{e}rn settings and 17 out of 300 across the three Egg Crate settings failed to converge. All replicates converged for our proposed method across all settings, for all parameter combinations. In the Egg Crate simulation, \texttt{SOAP} identified only one eigenfunction as optimal; therefore, the RMSE values for the second and third eigenfunctions are reported as NA. Lower values indicate better accuracy, and the values in bold represent the lowest median value for the corresponding simulation scenario. The proposed \texttt{mGSFPCA} demonstrates strong performance in the estimation of the eigenfunctions, achieving the lowest median \(\text{RMSE}_{\phi_k}\) in 13 out of the 18 scenarios presented. Notably, \texttt{mGSFPCA} shows improved performance in all scenarios with an Egg Crate covariance structure. The \texttt{CG} method was highly competitive, achieving the lowest median error in 6 of the 18 scenarios, all under the Mat\'{e}rn covariance. Overall, the three methods \texttt{mGSFPCA}, \texttt{CG}, and \texttt{ReMLE} demonstrated similar performances across most scenarios. As expected, all methods show improved estimates as the sample size increases (from Setting 1 to Setting 3).

\begin{table}[htp!]
\centering
\caption{Median (Interquartile Range) of $100\times$ the Root Mean Squared Error of the Top 3 Principal Eigenfunctions for each of the three simulated scenarios for both covariance structures. The smallest value in each column is highlighted in bold. \label{table:rmse_phi}}
\resizebox{\textwidth}{!}{
 \begin{tabular}{lccccccc}
 \hline
 & \multicolumn{3}{c}{\textbf{Mat\'{e}rn}} & & \multicolumn{3}{c}{\textbf{Egg Crate}} \\
 \cline{2-4}
 \cline{6-8}
  \multicolumn{8}{c}{\(\text{RMSE}_{\phi_1}\) Median(IQR) \(\times 100\)}\\
\hline
\textbf{Method}&Settings 1 & Settings 2 & Settings 3 & & Settings 1 & Settings 2 & Settings 3\\
\hline
\texttt{mGSFPCA} & 1.09  (0.73)  &  \textbf{0.76  (0.54)}  &  \textbf{0.55  (0.35)}  &&  \textbf{2.40  (2.00)}  &  \textbf{1.54  (1.18)}  &  \textbf{0.64  (0.42)}\\
\texttt{ReMLE} & 1.16  (0.69) & 0.82 (0.61) & 0.56  (0.29) & & 2.82  (1.54) & 1.96 (1.15) & 0.86  (0.44)\\
\texttt{CG} & \textbf{1.05 (0.68)} & 0.78 (0.58) & \textbf{0.55 (0.30)} && 2.76 (1.69) & 2.00 (1.04) & 0.78 (0.36) \\
\texttt{FACE1} & 2.35  (1.45) & 1.83 (1.37) & 1.01  (0.65) & & 3.18  (1.63) & 2.59 (1.37) & 1.56  (0.71)\\
\texttt{FACE2} & 1.36  (0.96) & 0.94 (0.52) & 0.62  (0.30) & & 2.73  (1.46) & 1.89 (0.92) & 0.84  (0.35)\\
\texttt{loc} & 2.48  (1.68) & 2.07 (1.34) & 1.30  (0.75) & & 3.58  (1.71) & 2.67 (1.43) & 1.51  (0.75)\\
\texttt{RKHS} & 2.42  (1.33) & 1.84 (0.99) & 1.26  (0.67) & & 3.92  (2.52) & 2.84 (1.43) & 1.82  (0.74)\\
\texttt{SOAP} &3.45  (2.54) & 3.29 (1.98) & 3.15  (1.21) & & 9.40  (1.27) & 8.26 (0.67) & 6.15  (0.39)\\
 \hline
 \multicolumn{8}{c}{\(\text{RMSE}_{\phi_2}\) Median(IQR) \(\times 100\)}\\
\hline
\texttt{mGSFPCA} & \textbf{1.51  (0.95)}  &  1.24  (0.67)  &  \textbf{0.76  (0.41)}  &&  \textbf{3.46  (2.24)}  &  \textbf{2.31  (1.82)}  &  \textbf{0.88  (0.65)} \\
\texttt{ReMLE} & 1.64  (0.92) & 1.32 (0.73) & \textbf{0.76  (0.39)} & & 4.62  (2.55) & 3.25 (1.77) & 1.28  (0.57)\\
\texttt{CG} & 1.58 (0.87) & \textbf{1.21 (0.62)} & 0.77 (0.44) && 4.73 (1.78) & 3.54 (1.44) & 1.33 (0.56)\\
\texttt{FACE1} & 3.63  (2.42) & 2.87 (1.5) & 1.44  (0.98) & & 6.30  (3.63) & 4.77 (2.80) & 2.47  (1.10)\\
\texttt{FACE2} & 2.04  (1.31) & 1.36 (0.88) & 0.91  (0.51) & & 4.28  (1.85) & 3.20 (1.80) & 1.24  (0.55)\\
\texttt{loc} & 4.51  (2.76) & 3.09 (1.79) & 1.89  (1.23) & & 7.35  (3.53) & 6.31 (2.34) & 4.78  (1.27)\\
\texttt{RKHS} & 3.99  (2.53) & 3.04 (1.61) & 2.11  (1.12) & & 7.04  (4.29) & 5.32 (2.40) & 3.23  (1.40)\\
\texttt{SOAP} &5.74  (2.24) & 5.02 (1.50) & 4.67  (0.71) & & NA  (NA) & NA (NA) & NA  (NA)\\
 \hline
 \multicolumn{8}{c}{\(\text{RMSE}_{\phi_3}\) Median(IQR) \(\times 100\)}\\
\hline
\texttt{mGSFPCA} & 2.18  (1.54)  &  1.78  (0.83)  &  1.18  (0.43)  &&  \textbf{2.31  (3.85)}  &  \textbf{2.69  (1.89)}  &  \textbf{0.94  (0.57)} \\
\texttt{ReMLE} & 2.19  (1.42) & 1.79 (0.91) & 1.20  (0.62) & & 6.94  (2.39) & 5.67 (2.29) & 1.46  (0.71)\\
\texttt{CG} & \textbf{2.11 (1.25)} & \textbf{1.69 (0.75)} & \textbf{1.06 (0.56)} && 6.91 (2.62) & 4.88 (2.42) & 1.59 (0.67) \\
\texttt{FACE1} & 4.16  (2.55) & 2.92 (1.62) & 2.28  (0.94) & & 9.64  (5.37) & 6.14 (3.68) & 3.98  (1.35)\\
\texttt{FACE2} & 2.32  (1.54) & 1.85 (0.94) & 1.26  (0.65) & & 6.40  (3.85) & 4.58 (2.21) & 2.02  (0.79)\\
\texttt{loc} & 5.25  (3.07) & 3.90 (2.48) & 3.19  (1.81) & & 9.80  (6.01) & 7.34 (4.43) & 5.28  (2.76)\\
\texttt{RKHS} & 4.60  (2.23) & 3.79 (1.57) & 3.26  (0.94) & & 11.81  (5.27) & 9.06 (4.29) & 4.86  (1.73)\\
\texttt{SOAP} &19.73  (0.00) & 18.07 (4.10) & 4.12  (3.23) & & NA  (NA) & NA (NA) & NA  (NA)\\
\hline
\end{tabular}   
}
\end{table}

Table \ref{table:se_lambda} presents the median and interquartile range (IQR) of \(10 \times \text{SE}_{\lambda_k}\) for the top three eigenvalue estimates (\(\lambda_1, \ \lambda_2\) and \(\lambda_3\)) across all simulation scenarios. A lower \(\text{SE}_{\lambda_k}\) value indicates greater accuracy in estimating these eigenvalues. The \texttt{SOAP} method is excluded from this table as it does not directly provide estimates for eigenvalues. Overall, the methods \texttt{mGSFPCA}, \texttt{CG}, and \texttt{ReMLE} showed broadly comparable performance across the simulation settings, with no differences observed in four scenarios. In particular, \texttt{mGSFPCA} consistently achieved the lowest values in all scenarios.

\begin{table}[htp!]
\centering
\caption{Median and Interquartile Range (IQR) of $10\times$ the Squared Error ($\text{SE}_{\lambda_k}$) of the Eigenvalues of the Top 3 Principal Eigenfunctions for each of the three simulated scenarios for both covariance structures. The smallest value in each column is highlighted in bold. Values reported as $0.00$ are less than $>0.001$.
\label{table:se_lambda}} 
\resizebox{\textwidth}{!}{
 \begin{tabular}{lccccccc}
 \hline
 & \multicolumn{3}{c}{\textbf{Mat\'{e}rn}} & & \multicolumn{3}{c}{\textbf{Egg Crate}} \\
 \cline{2-4}
 \cline{6-8}
  \multicolumn{8}{c}{\(\text{SE}_{\lambda_1}\) Median(IQR) \(\times 10\)}\\
\hline
\textbf{Method}&Settings 1 & Settings 2 & Settings 3 & & Settings 1 & Settings 2 & Settings 3\\
\hline
\texttt{mGSFPCA} &  \textbf{0.00  (0.01)}  &  \textbf{0.00  (0.00)}  &  \textbf{0.00  (0.00)}  &&  \textbf{0.00  (0.00)}  &  \textbf{0.00  (0.00)}  &  \textbf{0.00  (0.00)} \\
\texttt{ReMLE} & 3.68  (12.90) & 3.56 (7.39) & 3.23  (8.92) & & 0.06  (0.15) & 0.03 (0.08) & 0.01  (0.02)\\
\texttt{CG} & 4.58 (13.70) & 2.36 (7.23) & 1.75 (6.19) && 0.06 (0.12) & 0.03 (0.08) & 0.01 (0.02) \\
\texttt{FACE1} & 81.72  (179.71) & 37.78 (120.83) & 10.17  (31.07) & & 0.14  (0.36) & 0.11 (0.26) & 0.02  (0.06)\\
\texttt{FACE2} & 51.95  (102.78) & 18.80 (36.47) & 4.05  (12.52) & & 0.19  (0.35) & 0.05 (0.13) & 0.01  (0.03)\\
\texttt{loc} & 76.08  (184.90) & 37.02 (120.06) & 11.17  (28.00) & & 0.70  (0.86) & 0.60 (0.84) & 0.64  (0.29)\\
\texttt{RKHS} & 35.26  (90.92) & 15.33 (49.68) & 6.91  (23.16) & & 0.12  (0.31) & 0.08 (0.17) & 0.03  (0.06)\\
 \hline
 \multicolumn{8}{c}{\(\text{SE}_{\lambda_2}\) Median(IQR) \(\times 10\)}\\
\hline
\texttt{mGSFPCA} &  \textbf{0.00  (0.00)}  &  \textbf{0.00  (0.00)}  &  \textbf{0.00  (0.00)}  && \textbf{ 0.00  (0.00)}  & \textbf{ 0.02  (0.05)}  &  \textbf{0.00  (0.01)} \\
\texttt{ReMLE} & 2.42  (7.18) & 2.28 (8.11) & 2.20  (5.05) & & 0.02  (0.06) & \textbf{0.02 (0.05)} & \textbf{0.00  (0.01)}\\
\texttt{CG} & 3.30 (6.04) & 1.74 (7.29) & 1.49 (3.57) && 0.02 (0.04) & \textbf{0.02 (0.04)} & \textbf{0.00 (0.01)}\\
\texttt{FACE1} & 39.63  (80.57) & 22.01 (56.72) & 6.65  (18.68) & & 0.25  (0.43) & 0.14 (0.30) & 0.03  (0.07)\\
\texttt{FACE2} & 25.80  (44.36) & 7.85 (19.93) & 2.25  (5.69) & & 0.14  (0.18) & 0.05 (0.17) & 0.03  (0.03)\\
\texttt{loc} & 52.17  (123.67) & 27.65 (65.11) & 9.85  (33.62) & & 0.85  (0.54) & 0.99 (0.49) & 1.00  (0.23)\\
\texttt{RKHS} & 42.06  (98.84) & 16.80 (44.60) & 6.68  (16.81) & & 0.19  (0.47) & 0.08 (0.31) & 0.02  (0.05)\\
 \hline
 \multicolumn{8}{c}{\(\text{SE}_{\lambda_3}\) Median(IQR) \(\times 10\)}\\
\hline
\texttt{mGSFPCA} & \textbf{0.00  (0.00)}  &  \textbf{0.00  (0.00)}  &  \textbf{0.00  (0.00)}  && \textbf{0.00  (0.00)}  &  \textbf{0.01  (0.02)}  &  \textbf{0.00  (0.00)} \\
\texttt{ReMLE} & 1.36  (3.80) & 1.15 (3.59) & 0.68  (2.43) & & 0.01  (0.05) &\textbf{ 0.01 (0.02)} & \textbf{0.00  (0.00)}\\
\texttt{CG} & 0.95 (3.18) & 0.87 (2.42) & 0.50 (1.18) && 0.01 (0.03) & \textbf{0.01 (0.02)} & \textbf{0.00 (0.00)}\\
\texttt{FACE1} & 45.57  (90.90) & 30.27 (57.58) & 16.37  (27.16) & & 0.05  (0.13) & 0.02 (0.06) & 0.01  (0.02)\\
\texttt{FACE2} & 3.30  (12.82) & 3.33 (11.72) & 1.37  (2.79) & & 0.01  (0.03) & \textbf{0.01 (0.02)} & \textbf{0.00  (0.00)}\\
\texttt{loc} & 22.03  (46.34) & 14.30 (34.34) & 15.93  (22.10) & & 0.21  (0.19) & 0.23 (0.12) & 0.24  (0.08)\\
\texttt{RKHS} & 25.20  (44.55) & 27.26 (45.16) & 18.83  (39.27) & & 0.10  (0.15) & 0.06 (0.12) & 0.02  (0.03)\\
\hline
\end{tabular}   
}
\end{table}

\subsection{The Covariance Function, Error Variance, and Curve Reconstruction}
We assessed the accuracy of the estimated covariance function and error variance using two metrics: the RMSE for the covariance function, defined as \(\text{RMSE}_{\Sigma} = \sqrt{\frac{1}{50^2} \sum_{u=1}^{50} \sum_{j=1}^{50} (\hat{\Sigma}_{uj} - \Sigma_{uj})^2}\), and the SE for the error variance, given by \(SE_{\sigma^2} = (\hat{\sigma}^2 - \sigma^2)^2\).

Table \ref{table:rmse_C} summarizes the performance of our proposed method against the alternative approaches across the six simulation scenarios. For each method and scenario, we report the median and IQR of both $\text{RMSE}_{\Sigma}$ and \(SE_{\sigma^2}\). The \texttt{SOAP} method was excluded from covariance comparisons as it does not directly estimate eigenvalues. As shown in the upper section of Table  \ref{table:rmse_C}, \texttt{mGSFPCA} generally demonstrates improved performance for both the Mat\'{e}rn and Egg Crate covariance structures in terms of median $\text{RMSE}_{\Sigma}$. It achieved the lowest $\text{RMSE}_{\Sigma}$ in 4 of the 6 scenarios and ranks a close second to \texttt{ReMLE} in the remaining cases. The \texttt{CG} method exhibits higher median covariance estimation error relative to \texttt{mGSFPCA} across all six scenarios.

The lower section of Table \ref{table:rmse_C} presents the median and IQR of \(100 \times SE_{\sigma^2}\) for the error variance. The \texttt{RKHS} method does not provide an estimate of the error variance and is therefore not included. The \texttt{ReMLE} approach achieved the highest accuracy in all simulations based on a Matérn covariance structure. For the Egg-Crate simulation, \texttt{FACE2} attained the lowest value in Settings 1 and 2, whereas \texttt{ReMLE} and \texttt{mGSFPCA} achieved the lowest value in Setting 3.

\begin{table}[htp!]
\centering
\caption{Median and Interquartile Range (IQR) of the Root Mean Squared Error of the Covariance Function, $\text{RMSE}_{\Sigma},$ for each of the three simulated scenarios for the Mat\'{e}rn and Egg Crate covariances. The smallest value in each column is highlighted in bold.
\label{table:rmse_C}} 
\resizebox{\textwidth}{!}{
 \begin{tabular}{lccccccc}
 \hline
 & \multicolumn{3}{c}{\textbf{Mat\'{e}rn}} & & \multicolumn{3}{c}{\textbf{Egg Crate}} \\
 \cline{2-4}
 \cline{6-8}
  \multicolumn{8}{c}{\(RMSE_\Sigma\) Median(IQR)}\\
\hline
\textbf{Method}&Settings 1 & Settings 2 & Settings 3 & & Settings 1 & Settings 2 & Settings 3\\
\hline
\texttt{mGSFPCA} & 3.36  (1.16)  &  \textbf{2.68  (0.75)}  &  1.92  (0.58)  &&  \textbf{0.27  (0.09)}  &  \textbf{0.18  (0.07)}  &  \textbf{0.07  (0.02)} \\
\texttt{ReMLE} & \textbf{3.16  (1.21)} & 2.85 (2.85) & \textbf{1.81  (0.59)} & & 0.34  (0.12) & 0.27 (0.27) & 0.10  (0.02)\\
\texttt{CG} & 3.62 (1.00) & 3.26 (0.59) & 2.66 (0.39) && 0.33 (0.09) & 0.26 (0.07) & 0.10 (0.03) \\
\texttt{FACE1} & 7.90 (2.90) & 6.42 (6.42) & 3.73  (1.18) & & 0.44  (0.15) & 0.37 (0.37) & 0.22  (0.05)\\
\texttt{FACE2} & 4.71  (2.33) & 3.39 (3.39) & 2.05  (0.61) & & 0.35  (0.08) & 0.28 (0.28) & 0.13  (0.03)\\
\texttt{loc} & 8.07  (2.78) & 6.67 (6.67) & 4.62  (1.32) & & 0.56  (0.13) & 0.53 (0.53) & 0.47  (0.04)\\
\texttt{RKHS} & 8.38  (2.84) & 6.70 (6.70) & 4.71  (1.31) & & 0.51  (0.16) & 0.39 (0.39) & 0.24  (0.05)\\
 \hline
  \multicolumn{8}{c}{\(SE_{\sigma^2}\) Median(IQR) \(\times 100\)}\\
\hline
\texttt{mGSFPCA} & 59.80  (93.52)  &  11.80  (14.65)  &  3.32  (3.52)  && 0.98  (1.81)  &  0.27  (0.62)  &  \textbf{0.01  (0.02)} \\
\texttt{ReMLE} & \textbf{7.41  (12.72)} & \textbf{0.61 (1.97)} &\textbf{ 0.28  (14.74)} & & 0.70  (1.53) & 0.27 (0.60) & \textbf{0.01  (0.04)}\\
\texttt{FACE1} & 2846.90  (4778.61) & 1676.30 (3663.18) & 814.57  (1050.14) & & 6.46  (16.11) & 3.95 (8.19) & 1.10  (2.04)\\
\texttt{FACE2} & 62.39  (178.73) & 33.16 (89.29) & 18.17  (34.51) & & \textbf{0.39  (1.17)} & \textbf{0.13 (0.48)} & 0.02  (0.07)\\
\texttt{loc} & 755.59  (2297.16) & 306.65 (1359.22) & 238.98  (481.54) & & 12.10  (16.12) & 10.36 (11.65) & 14.22  (8.23)\\
\texttt{SOAP} &7188.92  (3179.45) & 5792.29 (3324.96) & 1131.95  (1321.46) & & 1236.12  (360.20) & 866.10 (195.29) & 223.33  (34.21)\\
\hline
\end{tabular}   
}
\end{table}

Finally, we evaluate the accuracy of the curve prediction on the evaluation grid \(\mathbf{g}\), using the RMSE, defined as \(\text{RMSE}_{X} = \sqrt{\frac{1}{50} \sum_{i=1}^{n} \sum_{j=1}^{50} (\hat{X}_i(g_j) - X_i(g_j))^2}\). Here, $\hat X_i$ is obtained by substituting the estimated eigenvalues $\hat{\boldsymbol{\lambda}}$, eigenfunctions $\hat{\boldsymbol{\phi}}$, and mean function $\hat{\boldsymbol{\mu}}$ into the conditional expectation formula in~(\ref{CE}). An exception is the \texttt{SOAP} method, which directly estimates the trajectories without relying on the reconstruction in~(\ref{Aprox_X}). The \texttt{RKHS} and \texttt{CG} methods are excluded from this analysis as they do not provide the mean and/or the error variance estimates required for computing (\ref{Aprox_X}) and (\ref{CE}). Table \ref{table:rmse_XHat} presents the median and IQR of $\text{RMSE}_{X}$ for each of the six scenarios. 

The curve prediction results indicate that \texttt{ReMLE} achieved the highest overall accuracy, yielding the most accurate estimates in four of the six simulation settings, followed closely by \texttt{mGSFPCA} and \texttt{FACE2}.
Curve prediction accuracy is sensitive to the quality of the error variance estimate $\hat{\sigma}^2$, particularly because of the matrix inversion involved in~(\ref{Aprox_X}). This sensitivity likely accounts for the slightly reduced accuracy observed for \texttt{FACE1} and \texttt{loc}, both of which exhibited less accurate error variance estimates in Table (\ref{table:rmse_C}). 

\begin{table}[htp!]
\centering
\caption{Median and Interquartile Range (IQR) of the RMSE of the curve predictions ($\text{RMSE}_{\hat{X}}$) across the three scenarios for the Mat\'{e}rn and Egg Crate covariance structures. The smallest value in each column is highlighted in bold.} 
\label{table:rmse_XHat}
\resizebox{\textwidth}{!}{
 \begin{tabular}{lccccccc}
 \hline
 \multicolumn{8}{c}{\(\text{RMSE}_{X}\) Median(IQR)}\\
\hline
 & \multicolumn{3}{c}{\textbf{Mat\'{e}rn}} & & \multicolumn{3}{c}{\textbf{Egg Crate}} \\
 \cline{2-4}
 \cline{6-8}
\textbf{Method}&Settings 1 & Settings 2 & Settings 3 & & Settings 1 & Settings 2 & Settings 3\\
\hline
\texttt{mGSFPCA} &  1.91  (0.33)  &  1.96  (0.29)  &  2.69  (0.13)  &&  \textbf{0.65  (0.12)}  & \textbf{ 0.60  (0.04)}  &  \textbf{0.53  (0.03)} \\
\texttt{ReMLE} & \textbf{1.75  (0.37)} & \textbf{1.87 (0.21)} & \textbf{2.63  (0.10)} & & 0.68  (0.07) & 0.63 (0.05) & \textbf{0.53  (0.03)}\\
\texttt{FACE1} & 2.32  (0.61) & 2.23 (0.47) & 2.85  (0.19) & & 0.72  (0.07) & 0.67 (0.05) & 0.59  (0.03)\\
\texttt{FACE2} &\textbf{ 1.75  (0.30)} & \textbf{1.87 (0.22)} & 2.64  (0.13) & & 0.68  (0.07) & 0.64 (0.06) & 0.55  (0.03)\\
\texttt{loc} & 2.28  (0.61) & 2.26 (0.78) & 2.97  (5.78) & & 0.77  (0.08) & 0.72 (0.06) & 0.70  (0.03)\\
\texttt{SOAP} &4.22  (0.29) & 4.23 (0.34) & 4.46  (0.38) & & 2.46  (0.12) & 2.26 (0.05) & 1.91  (0.05)\\
\hline
\end{tabular}  
}
\end{table}

\subsection{Model Selection}\label{sec:model_sel}

\begin{figure}[htp]
    \centering
    \includegraphics[width=0.80\textwidth]{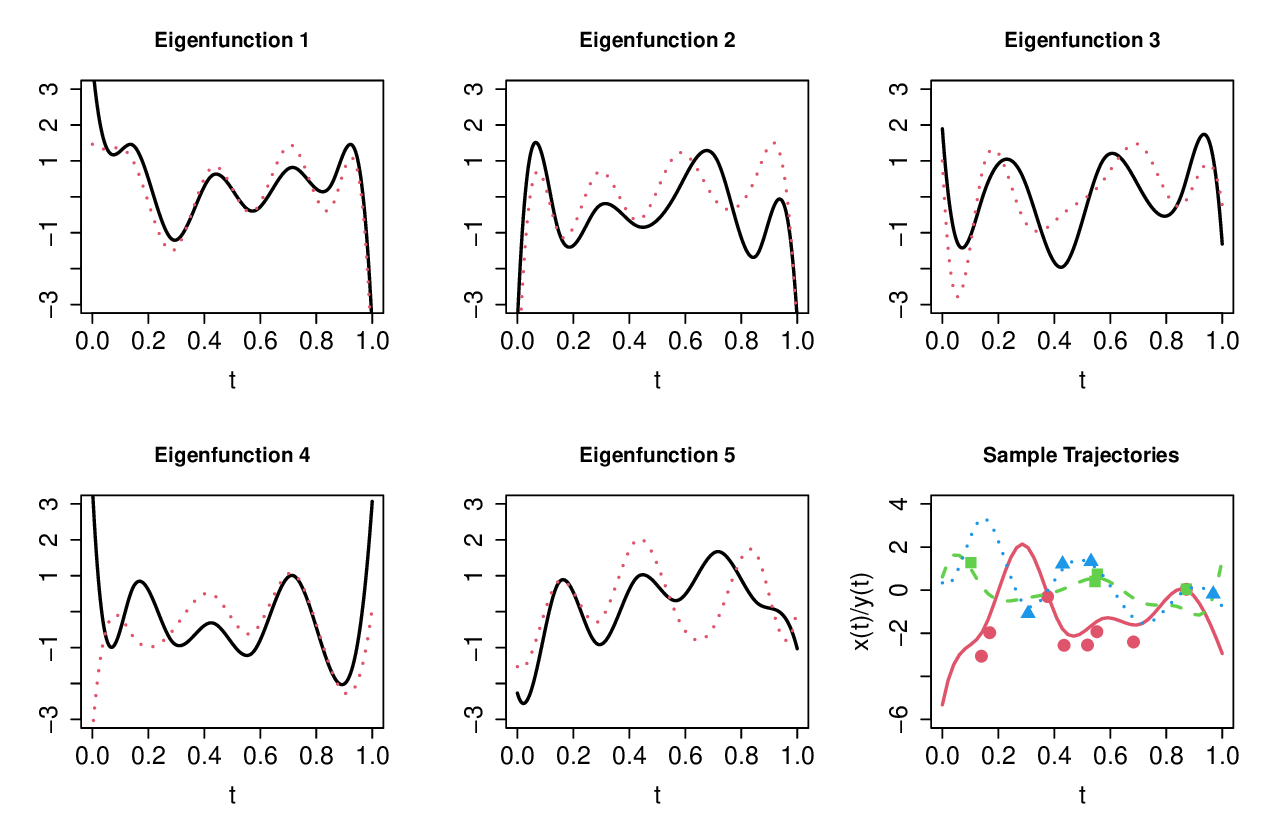}
    \caption{True (solid lines) and estimated (dotted lines) principal eigenfunctions and sample trajectories for the cubic b-spline simulation.
}
    \label{fig:sim2}
\end{figure}

To investigate the performance of model selection using the AIC-type criterion, defined in (\ref{eqn:GCV}), we conducted simulation studies similar to those used in \cite{Peng2009}. In these simulations, data were generated with known eigenvalues and eigenfunctions, represented using cubic B-splines. For the first simulation, we set the eigenvalues as $\{\lambda_k\}_{k=1}^5 = k^{-0.6}$, with the true eigenfunctions represented by \(Q = 10\) cubic B-spline basis functions with equally spaced knots. Data with a sample size of $n = 100$ were generated using principal component scores $\xi_i \sim N(0,1)$ with error variance $\sigma^2 = 1/16$. The simulation was replicated 100 times. Figure \ref{fig:sim2} provides an illustrative example of these eigenfunctions. The second simulation utilized three leading eigenvalues $(1, 0.66, 0.52)$, a fourth eigenvalue ($0.07$) comparable to the error variance, and six additional smaller eigenvalues $(9.47 \times 10^{-3}, 1.28 \times 10^{-3}, 1.74 \times 10^{-4}, 2.35 \times 10^{-5}, 3.18 \times 10^{-6}, 4.30 \times 10^{-7})$. In this setup, the first three eigenvalues account for 96.4\% of the total signal variability, while including the fourth eigenvalue explains 99.5\%. Data were generated with a sample size of $n = 500$ and error variance $\sigma^2 = 1/16$.

Simulation results concerning the accuracy of the estimated eigenfunctions and eigenvalues, similar to those presented in Table \ref{table:rmse_phi} and Table \ref{table:se_lambda} for the previous simulation, are provided in Appendix A.5.3. The focus here is on evaluating the selection of the optimal number of basis functions ($Q$) and the optimal rank ($p$).

In addition to the proposed AIC-type criterion in~\eqref{eqn:GCV}, we evaluate parameter selection using $K$-fold cross-validation (CV) with $K = 5$ and $K = 10$. The $K$-fold CV score is defined as
\[
  \mathrm{CV} = \sum_{k=1}^{K} \sum_{i \in \mathcal{S}_k} 
  \mathcal{L}_i\!\bigl(\mathbf{Y}_i, \mathbf{t}_i, 
  \hat{\boldsymbol{\Theta}}^{(-\mathcal{S}_k)}\bigr),
\]
where $\mathcal{S}_k$ denotes the set of curve indices in the $k$-th fold and $\hat{\boldsymbol{\Theta}}^{(-\mathcal{S}_k)} = (\hat{\mathbf{C}}, \hat{\boldsymbol{\eta}}, \hat{\gamma})^{(-\mathcal{S}_k)}$ is the parameter estimate obtained from the data excluding the $k$-th fold. This CV score corresponds to the empirical predictive Kullback--Leibler risk, which, as noted by~\citet{Peng2009}, is more appropriate than prediction error loss for the current problem.

The first simulation focused on evaluating the selection of the optimal number of basis functions ($Q$) given a true rank ($p$). Under this setting, \texttt{mGSFPCA} successfully selected the true value of $Q=10$ in 96 out of the 100 replicates, when considering possible models with $Q \in \{5, 10, 15, 20\}$. All 4 possible models for each replicate for this simulation converged successfully. The model with $Q = 15$ was selected 4 times. For the 5-fold CV, $Q = 10$ was selected in 94 out of 100 replicates, with $Q = 5$ and $Q = 15$ each selected 3 times. The 10-fold CV gave similar results, selecting $Q = 10$ in 93 replicates, $Q = 5$ in 2, and $Q = 15$ in 5.

The second simulation aimed to assess the simultaneous selection of both the number of principal components ($p$) and the number of basis functions ($Q$). We fit models considering \(Q \in \{8, 9, \ldots, 15\}\) basis functions and \(p \in \{2, \ldots, 6\}\) principal components. We explore two distinct model selection approaches for \texttt{mGSFPCA}. The first, denoted as \texttt{\(\text{mGSFPCA}_{Q\times p}\)}, involves selecting the optimal model from a grid of $Q$ and $p$ parameter combinations. The second approach, \texttt{\(\text{mGSFPCA}_{Q + p}\)}, employs a sequential selection process which would be computationally more efficient: first, $Q$ is determined by setting $p$ to its maximum considered value, and then, $p$ is selected using the previously determined $Q$. Note that the sequential approach requires \(max(p) \leq min(Q)\). 


For both the \texttt{mGSFPCA$_{Q \times p}$} and \texttt{mGSFPCA$_{Q + p}$} selection strategies, the AIC correctly identified the true number of basis functions ($Q = 10$) in 100\% of the replicates. With respect to the number of principal components ($p$), the AIC showed consistent performance across both approaches. Specifically, it selected $p = 4$ in 53 out of 100 replicates, corresponding to a fraction of explained variance (FEV) of 0.995 in the true simulated data. In 45 replicates, $p = 5$ was selected (FEV $\approx 0.999$), and in the remaining 2 replicates, $p = 6$ was selected ( FEV $\approx0.9999$).

Comparing the proposed AIC-type criterion to CV, both approaches exhibit comparable accuracy in identifying the true $Q=10$, with 10-fold CV and 5-fold CV selecting the correct number of basis functions in 99\% and 100\% of replicates, respectively. The two CV variants themselves yield nearly identical results across all settings, suggesting that the more expensive 10-fold CV provides no practical advantage over the 5-fold CV. However, both CV schemes tend to overestimate the number of principal components, predominantly selecting $p=6$ in 80 and 76 replicates for 10-fold and 5-fold CV, respectively. 

We observe that the \texttt{mGSFPCA$_{Q \times p}$} and \texttt{mGSFPCA$_{Q + p}$} approaches demonstrate performance highly similar to each other and to CV. In some cases, they outperform CV by yielding more parsimonious models. In addition, the \texttt{mGSFPCA$_{Q + p}$} approach achieves comparable results while requiring the evaluation of only 13 candidate models, a substantial reduction from the 40 models considered by \texttt{mGSFPCA$_{Q \times p}$}, leading to gains in computational efficiency.

\subsection{Computational Performance}\label{sec:comp}

Table \ref{table:time} summarizes the median and interquartile range of computation times in seconds for each method. These timings were obtained using the simulation settings in Section \ref{sec:model_sel}, with \(n = 100\) and \(n = 500\) samples. For \texttt{mGSFPCA}, \texttt{ReMLE}, and \texttt{CG}, we considered models with \(Q = \) 5, 10, 15, 20 and \(p = \) 3, 4, 5, for their computations. All simulations were performed using R version 4.3.1 on an Intel Core i9 computer with 32GB RAM.

The proposed \texttt{mGSFPCA} demonstrates competitive computational efficiency compared to existing approaches, with median times ranging from 5.40 to 15.56 seconds across sample sizes. As expected, methods with closed-form solutions (\texttt{FACE2} and \texttt{loc}) generally show faster computation. It should be noted that the high computation time for \texttt{loc} at \(n=500\) is primarily attributed to calculating the conditional expectation rather than the covariance estimation. Among methods requiring numerical optimization (\texttt{mGSFPCA}, \texttt{ReMLE}, \texttt{CG}, and \texttt{RKHS}), the proposed \texttt{mGSFPCA} consistently shows improved performance for both sample sizes. To ensure a fair comparison, the \texttt{RKHS} method must be run through both the estimation stage and the FPCA evaluation stage.

\begin{table}[htp!]
\centering
\caption{Median and Interquartile range (IQR) of computation times (in seconds) for each method.} 
\label{table:time}
\resizebox{\textwidth}{!}{
 \begin{tabular}{ccccccccc}
 \hline
  \multicolumn{8}{c}{\(\text{Median (IQR) Computation Time }\)} \\
\hline
 \textbf{n}& \texttt{\(\text{mGSFPCA}_{Q\times p}\)} & \texttt{\(\text{mGSFPCA}_{Q + p}\)} & \texttt{ReMLE} & \texttt{FACE2} & \texttt{loc} & \texttt{RKHS} & \texttt{CG}\\
\hline
100 & 6.73 (3.76) & 5.40 (3.88) & 23.30 (2.72) & 2.20 (0.02) & 0.57 (0.15) & 38.60 (20.31) & 100.04 (15.05)\\
500 & 15.56 (5.01) & 11.78 (4.28) & 90.69 (8.47) & 9.44 (0.15) & 27.62 (4.54) & 160.56 (8.81) & 227.82 (34.20)\\
\hline
\end{tabular}   
}
\end{table}

\section{Applications}\label{apps}

\subsection{CD4 Cell Counts}

CD4 cells are essential components of the human immune system, playing a pivotal role in activating and coordinating responses to infections and diseases. These cells enhance the ability of B cells to produce antibodies targeting specific pathogens and support the activation and proliferation of cytotoxic T cells, crucial for eliminating infected or diseased cells. In the context of HIV infection, the virus specifically targets and destroys CD4 cells, leading to a weakened immune system and heightened susceptibility to various infections and diseases. Consequently, monitoring CD4 cell counts is critical for managing HIV infection and assessing the effectiveness of antiretroviral therapy.

The objectives of our analysis are to estimate the overall trend in CD4 cell counts over time, extract the dominant modes of variation, and recover individual trajectories from sparse measurements. We apply our proposed method to CD4 cell count data from the Multicenter AIDS Cohort Study \citep{Kaslow1987}. The specific version of the data used in this analysis is obtained from the \texttt{refund} R package \citep{Goldsmith2010}.
The dataset comprises 1,881 longitudinal measurements of CD4 cell counts from 366 HIV-infected men. The number of observations per subject ranges from 1 to 11, with a median of 5. Spanning approximately 8.5 years, the study period includes about 3 years before and 5.5 years after seroconversion, the point at which HIV becomes detectable. This dataset has been previously analyzed by  \cite{Diggle2002}, \cite{Yao2005}, \cite{Peng2009} and \cite{Xiao2018}. The data is skewed; therefore, we apply a square root transformation to the CD4 cell counts. We used B-spline basis functions, with the number of basis functions and the number of principal components determined using (\ref{eqn:GCV}). This approach selected 3 principal components from \(\{3, \cdots, 5\}\) and 5 basis functions from \(\{5, \cdots, 11\}\). 

\begin{figure}[htp!]
    \centering
    \includegraphics[width=.85\textwidth]{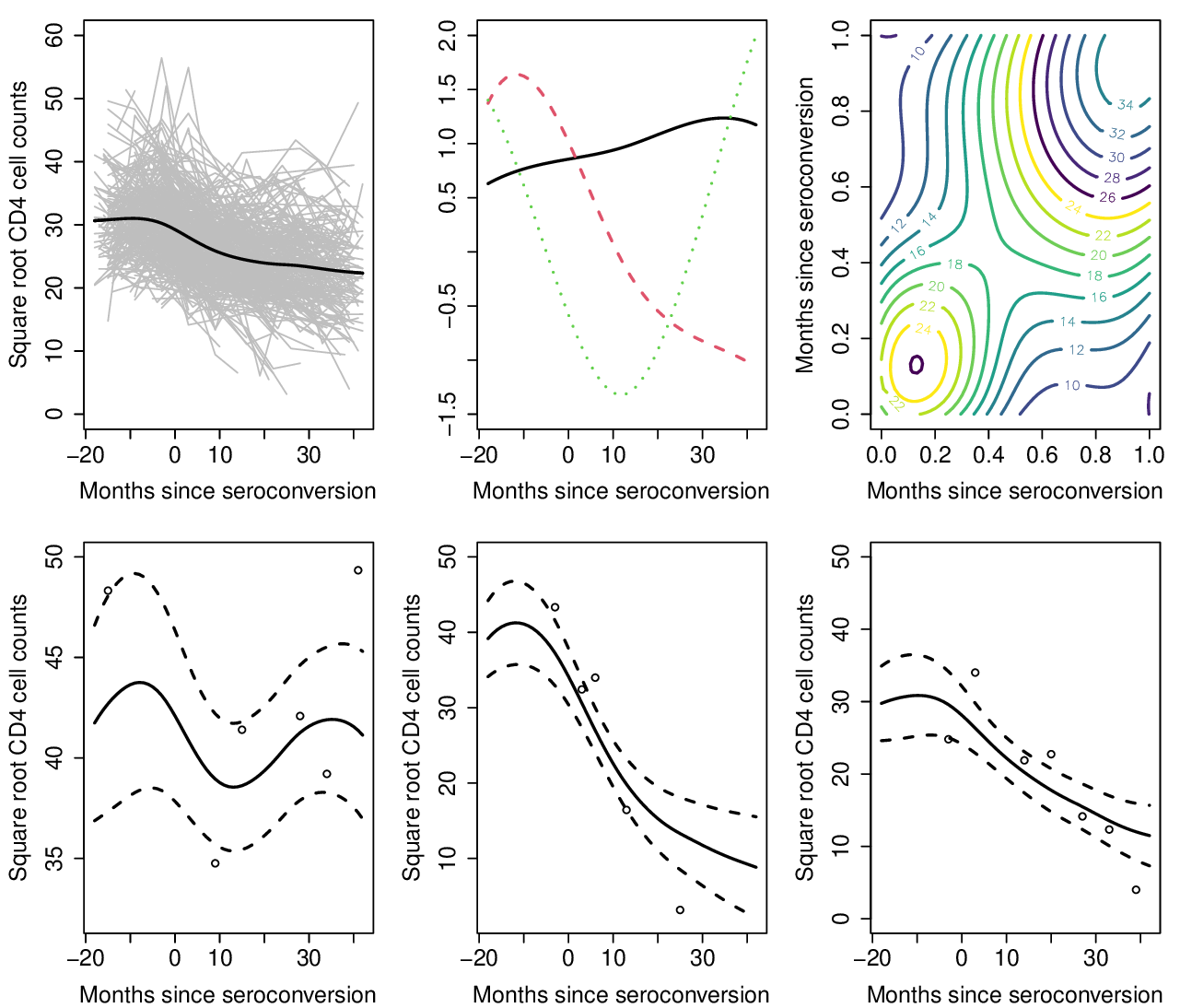}
    \caption{Estimates of the mean, eigenfunctions, covariance surface, and sample subject trajectories of the CD4 cell count data. Top row left: Estimated mean (thick dark line) and sample trajectories (thin gray lines); Top row middle: Estimated eigenfunctions: \(\hat{\phi}_1\) (solid line), \(\hat{\phi}_2\) (dashed line), and \(\hat{\phi}_3\) (dotted line); Top row right: Estimated covariance contour plot; Bottom row panels: Observations (circles), predicted trajectories (solid line), and 95\% point-wise confidence bands (dashed lines) for selected subjects. These subjects correspond to those with the largest absolute scores for each of the three principal components respectively.}
    \label{fig:cd4_eigen}
\end{figure} 

Figure \ref{fig:cd4_eigen} displays estimates of the mean function, eigenfunctions, covariance surface, and predicted trajectories for selected individuals. The first principal component (solid line in the top-middle panel), accounting for approximately 77\% of the total data variation, appears flat and captures inter-subject shifts from the mean. The predicted trajectory for the individual with the highest absolute principal score corresponding to this component is shown in the bottom-left panel, representing the subject with the largest vertical deviation from the mean. The second principal component (dashed line in the top-middle panel) accounts for approximately 21\% of the total variation. Its shape resembles the mean function, indicating a diminishing trend in CD4 cell counts over time. The subject with the highest absolute score for this component is presented in the bottom-middle panel; the similarity between their predicted trajectory and the principal component suggests a more rapid decrease in CD4 cell count for this individual.  Finally, the third principal component (dotted line in the top-middle panel) accounts for only approximately 2\% of the total variation. The subject with the highest absolute score for this component is depicted in the bottom-right panel. Since this component explains a very small proportion (2\%) of the total variation, any interpretation of its trajectory should be treated with caution.

\subsection{Somatic Cell Counts}

Somatic cells, naturally present in milk, are white blood cells that play a crucial role in fighting infection and repairing tissue damage. When a cow's udder is infected, these cells migrate into the milk to combat the invading bacteria. Thus, the somatic cell count (SCC) of a cow's milk is a key indicator of udder health and milk quality. The objectives of our analysis are to estimate the overall trend in SCC in cows' milk over time, extract dominant modes of variation, and determine how individual cows' SCC levels deviate from the overall trend. We applied our proposed method to data representing the average weekly morning somatic cell score (SCS), which is $\log_2$(SCC), from 282 cows across various research farms in Ireland, covering a 44-week milking period that began five days post-calving for the year 2018.

We employed B-spline basis functions, with the number of basis functions and the number of principal components determined using the criterion in (\ref{eqn:GCV}). This approach selected five principal components and seven basis functions from the possible models with \(p \in \{3, \cdots, 6\} \) and \(Q = \{5, \cdots 11\}\). The first three principal components account for 94\% of the total variation in the functional data: the first component explains 80\%, the second 9\%, and the third 5\%.

\begin{figure}[!htp]
    \centering
    \includegraphics[width=0.85\textwidth]{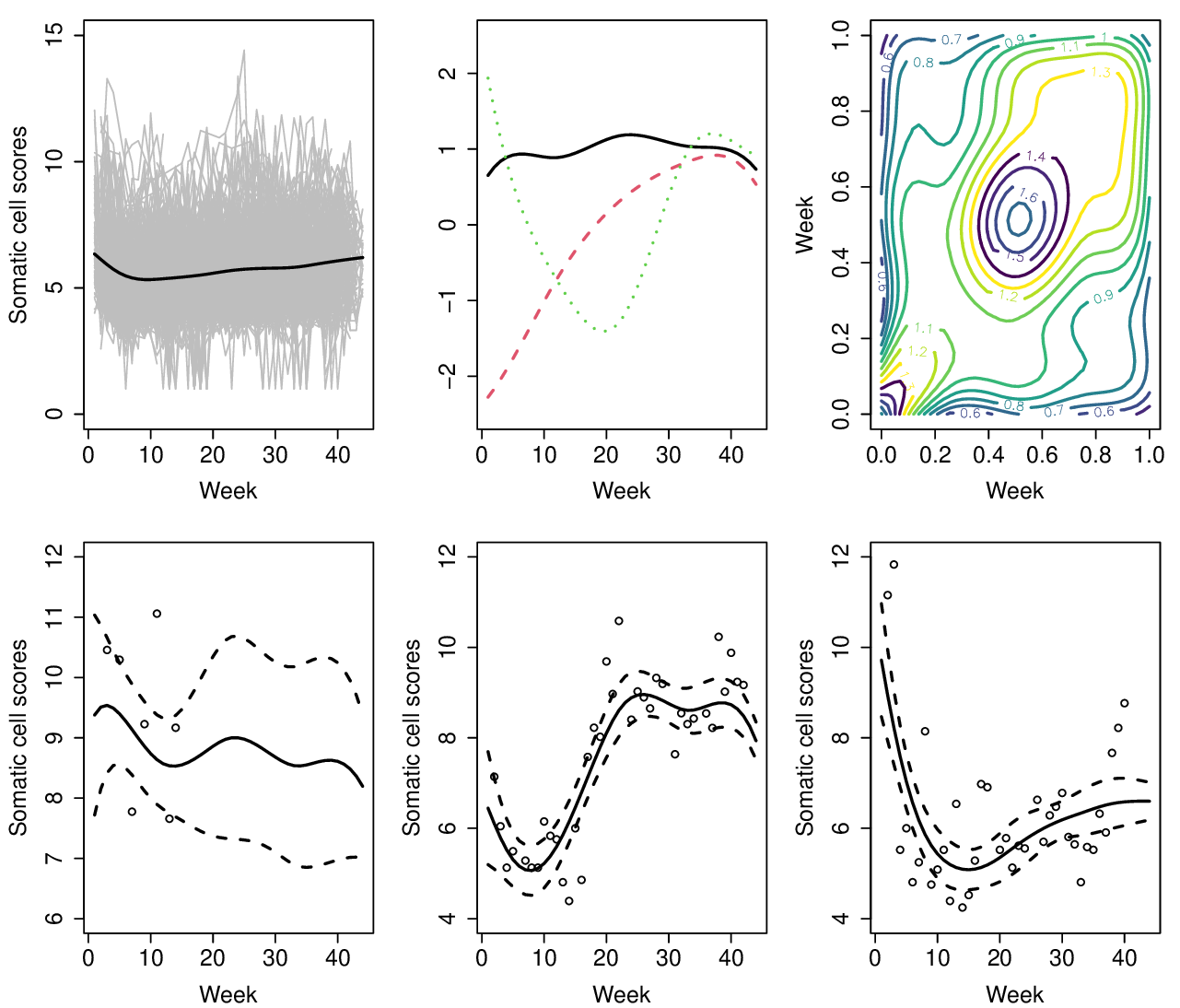}
    \caption{Estimates of the mean, eigenfunctions, covariance surface, and sample subject trajectories of the Somatic cell score data. Top row left: Estimated mean (thick dark line) and sample trajectories (thin gray lines); Top row middle: Estimated eigenfunctions: \(\hat{\phi}_1\) (solid line), \(\hat{\phi}_2\) (dashed line), and \(\hat{\phi}_3\) (dotted line); Top row right: Estimated covariance contour plot; Bottom row panels: Observations (circles), predicted trajectories (solid line), and 95\% point-wise confidence bands (dashed lines) for selected subjects. These subjects correspond to those with the largest absolute scores for each of the three principal components, respectively.}
    \label{fig:scc_eigen}
\end{figure}

Figure \ref{fig:scc_eigen} presents estimates of the mean function, eigenfunctions, covariance surface, and predicted trajectories for selected individuals.
The top-middle panel of Figure \ref{fig:scc_eigen} displays the estimated top three eigenfunctions. The first component (solid line) remains relatively flat over time, primarily capturing baseline variability in SCS across individual cows. The second eigenfunction (dashed line) appears to reflect a transition from low to high SCS values over time. The third eigenfunction (dotted line) seems to characterize a recovery pattern in SCS, exhibiting an initial decline followed by an increase.
The bottom row of Figure \ref{fig:scc_eigen} shows the predicted trajectories (solid lines), observations (circles), and 95\% point-wise confidence bands (dashed lines) for cows exhibiting the largest absolute principal component scores in the direction of the top three eigenfunctions, respectively. The shapes of these predicted trajectories closely align with their corresponding eigenfunctions, providing visual confirmation of the principal components' interpretability.

\section{Conclusion}\label{sec:conc}
This paper presents a new method for functional principal component analysis (FPCA) designed for sparsely observed functional data. Our approach addresses a key limitation of existing methods that rely on optimization over the Stiefel manifold, which can lead to numerical instability in practice. By integrating a MGS orthonormalization process directly into the estimation procedure, the proposed method (\texttt{mGSFPCA}) avoids direct optimization of the eigenfunction coefficients on the Stiefel manifold. Although manifold-based methods have better theoretical properties, we find that, in finite-precision computations, the MGS formulation is more stable. The MGS orthonormalization ensures orthonormality of the eigenfunctions while enabling the use of standard quasi-Newton optimization for efficient parameter estimation. The proposed framework simultaneously estimates the eigenfunctions, eigenvalues, and error variance via maximum likelihood, while guaranteeing a positive estimate of the error variance.

Through extensive simulation studies, we demonstrated that the proposed \texttt{mGSFPCA} performs better or is comparable to several state-of-the-art methods in accurately estimating eigenfunctions, eigenvalues, covariance functions, and error variance. In particular, \texttt{mGSFPCA} showed very good performance in estimating the eigenfunctions and generally achieved low squared errors for eigenvalues across various settings and covariance structures. \textcolor{black}{While \texttt{CG} and \texttt{ReMLE} showed better accuracy in some simulation scenarios, \texttt{mGSFPCA} consistently converged across all simulation replicates, unlike \texttt{ReMLE} and \texttt{CG}. Moreover, the \texttt{mGSFPCA} approach demonstrates greater computational efficiency relative to methods requiring numerical optimization.}

Furthermore, the proposed AIC-type criterion proved effective in selecting the optimal number of basis functions (\(Q\)) and principal components (\(p\)). The sequential model selection approach (\texttt{\(\text{mGSFPCA}_{Q + p}\)}) also demonstrated significant computational efficiency gains by reducing the number of models evaluated compared to the grid search approach (\texttt{\(\text{mGSFPCA}_{Q\times p}\)}), while maintaining a highly comparable performance.

The practical utility of \texttt{mGSFPCA} was demonstrated through its application to the CD4 cell count dataset from the Multicenter AIDS Cohort Study and the somatic cell counts of cows from Irish research dairy farms. Our analysis effectively estimated the overall trend in CD4 cell counts and somatic cell counts over time and identified the dominant modes of variation, which collectively accounted for a significant portion of the total variance in the data. The ability to accurately reconstruct individual trajectories from sparse measurements highlights the method's potential for longitudinal studies and clinical monitoring.

Future research directions include extending \texttt{mGSFPCA} to handle multivariate functional data and exploring its applicability in contexts with more complex data structures, such as those involving functional snippets or non-Gaussian error distributions. An important theoretical question concerns the relationship between the MGS parameterization used here and the Stiefel manifold formulation of \cite{Peng2009}. Although both approaches optimize the same likelihood and are linked via (\ref{Ctilde}), establishing formal equivalence of the resulting estimators and their asymptotic properties requires a rigorous analysis that lies beyond the scope of this paper. We leave a full treatment of this equivalence to future work.

\section{Supplementary Materials}
\begin{description}
\item[Appendix:] The appendix contains extra details and simulation results (supp.pdf).
\item[R-package for mGSFPCA routine:] R-package \texttt{mGSFPCA} containing code to perform the estimation and model procedure described in the article. The package also contains an example dataset used in the simulation studies (mGSFPCA\_0.2.2.tar.gz)
\item[Example Simulation:] R script containing the basic implementation of all the simulations used in the study.
\end{description}

\section{Data Availability Statement }
The CD4 cell count data were derived from the following resources available in the public domain: \url{https://rdrr.io/cran/refund/man/cd4.html}. The Somatic Cell Count data that support the findings of this study are available from Teagasc. Restrictions apply to the availability of these data, which were used under license for this study. Data are available from the authors upon reasonable request with the permission of Teagasc. 

\section{Disclosure Statement}

The authors report that there are no competing interests to declare.

\section{Additional information}

\subsection{Acknowledgement}

We are grateful to the Associate Editor and the two referees for their insightful and helpful comments, which have greatly improved the quality of this work. We acknowledge the financial support of Research Ireland (SFI) and the Department of Agriculture, Food and Marine on behalf of the Government of Ireland under Grant Number [16/RC/3835] - VistaMilk and Research Ireland (SFI) and co-funding partners under grant number 21/SPP/3756 through the NexSys Strategic Partnership Programme. The authors would like to thank Blerta Begu for helpful discussions and insights that contributed to this work. The authors wish to also thank Dr. Emer Kennedy and Dr. Pablo Silva Bolona for their valuable insight in discussions on the somatic cell count variation in diary cattle. To improve the language and clarity of the manuscript, we used the generative AI tools DeepSeek-V3.2, ChatGPT 5.4, and Claude Sonnet 4.5.

\newpage
\bibliographystyle{apalike}
\bibliography{ref.bib}

\end{document}


\maketitle

\newpage

\section*{A.1. Direct Construction of Orthonormal Eigenfunctions}\label{app:Orth}

Following \cite{Peng2009} and \cite{He2022}, let the $k^{th}$ eigenfunction, $\phi_k(t)$, be modelled using a basis expansion: $\phi_k(t) = \sum_{l = 1}^Q \tilde{c}_{lk}\tilde{B}_{l}(t)$, where $\{\tilde{B}_{l}(\cdot)\}_{l=1}^Q$ (with $Q \geq p$) are orthonormal basis functions and $\boldsymbol{\tilde{C}} = ((\tilde{c}_{kl}))$ is a $Q \times p$ matrix of  orthonormal basis coefficients. Substituting the basis expansion in for the eigenfunctions results in
\begin{equation*}
    \begin{split}
        \textbf{I}_p &= \int_0^1 \boldsymbol{\Phi}(t)^T \boldsymbol{\Phi}(t) \textrm{dt},\\
        &= \int_0^1 (\boldsymbol{\tilde{B}}(t) \tilde{\textbf{C}})^T (\boldsymbol{\tilde{B}}(t)\tilde{\textbf{C}}) \textrm{dt}, \\ 
        &= \int_0^1 \tilde{\textbf{C}}^T \boldsymbol{\tilde{B}}(t)^T \boldsymbol{\tilde{B}}(t) \tilde{\textbf{C}} \textrm{dt}, \\
        & = \tilde{\textbf{C}}^T \left( \int_0^1 \boldsymbol{\tilde{B}}(t)^T \boldsymbol{\tilde{B}}(t) \textrm{dt} \right) \tilde{\textbf{C}}, \\
        & =  \tilde{\textbf{C}}^T \tilde{\textbf{C}}.
    \end{split}
\end{equation*}

To ensure $\textbf{I}_p = \int_0^1 \boldsymbol{\Phi}(t)^T \boldsymbol{\Phi}(t) \textrm{dt},$  \cite{Peng2009} and \cite{He2022} must estimate $\tilde{\textbf{C}}$ with the constraint that $\tilde{\textbf{C}}^T \tilde{\textbf{C}}=\textbf{I}_p.$ In contrast, the proposed approach enforces the orthonormality constraint, $\mathbf{I}_p = \int_0^1 \boldsymbol{\Phi}(t)^\top \boldsymbol{\Phi}(t)\, dt,$
directly by constructing an orthonormal system of eigenfunctions via a weighted modified Gram--Schmidt (MGS) procedure,
$
\boldsymbol{\Phi} = \mathcal{M}_W(\mathbf{B}\mathbf{C}),
$
where $\mathbf{B}$ denotes an unconstrained basis matrix, $\mathbf{C}$ is the corresponding coefficient matrix, and $\mathbf{W}$ is a diagonal matrix containing the quadrature weights.

\begin{proposition}[Grid-level orthonormality via weighted MGS]
\label{prop:MGS_tau_orth}
Let $\boldsymbol{\tau}=\{\tau_j\}_{j=1}^M\subset[0,1]$ be a grid with associated quadrature
weights $w_j>0$, and define the weighted inner product
\[
\langle f,g\rangle_w = \sum_{j=1}^M w_j f(\tau_j)g(\tau_j),
\qquad
\|f\|_w = \sqrt{\langle f,f\rangle_w}.
\]
Let $\mathbf U=[u_1(\boldsymbol{\tau}),\dots,u_p(\boldsymbol{\tau})]\in\mathbb{R}^{M\times p}$ be any
collection of linearly independent functions evaluated on $\boldsymbol{\tau}$.
If the MGS procedure is applied to $\mathbf U$
with respect to $\langle\cdot,\cdot\rangle_w$, then the resulting functions
$\boldsymbol{\Phi}=[\phi_1(\boldsymbol{\tau}),\dots,\phi_p(\boldsymbol{\tau})]$ satisfy
$\boldsymbol{\Phi}^\top \textbf{W} \boldsymbol{\Phi} = \textbf{I}_p$,
where $\textbf{W}=\mathrm{diag}(w_1,\dots,w_M)$. That is, the functions
$\{\phi_k\}_{k=1}^p$ are orthonormal on the grid $\boldsymbol{\tau}$ with respect to
the weighted discrete inner product.
\end{proposition}

\begin{proof}
The MGS procedure constructs $\phi_k$ recursively by
subtracting from $u_k$ its projections onto the previously computed
$\{\phi_\nu\}_{\nu<k}$ using the inner product $\langle\cdot,\cdot\rangle_w$,
followed by normalization with respect to the induced norm $\|\cdot\|_w$.
By the standard properties of Gram--Schmidt orthogonalization in an inner
product space see \cite{Golub2013} for details, the resulting functions satisfy
$
\langle \phi_k,\phi_\nu\rangle_w = \delta_{k\nu},$ for $k,\nu=1,\dots,p.$
In matrix form, this condition is equivalent to
$\boldsymbol{\Phi}^\top \textbf{W} \boldsymbol{\Phi}=\mathbf I_p$.
\end{proof}

\begin{remark}
In Peng and Paul (2009), orthonormality of the eigenfunctions is enforced by assuming an $L^2[0,1]$-orthonormal basis $\{\tilde{B}_{l}(\cdot)\}_{l=1}^Q$ and
imposing $\tilde{\mathbf C}^\top\tilde{\mathbf C}=\mathbf I_p$, which yields
$\int_0^1\boldsymbol{\Phi}(t)^\top\boldsymbol{\Phi}(t)\,dt=\mathbf I_p.$
In contrast, the present approach enforces orthonormality
directly on the grid $\boldsymbol{\tau}$ via weighted MGS. When the quadrature rule is
consistent, i.e.,
\[
\sum_{j=1}^M w_j f(t_j)g(t_j)
\;\to\;
\int_0^1 f(t)g(t)\,dt
\quad \text{as } M\to\infty,
\]
grid-level orthonormality converges to continuous $L^2[0,1]$ orthonormality and the two approaches are asymptotically equivalent.
\end{remark}

\section*{A.2. The Gradient of \texorpdfstring{$\mathcal{L}$}{Negative Log-Likelihood}}\label{appx:grad}

\subsection*{A.2.1. Coefficients}
The derivative of the average negative log-likelihood loss $\mathcal{L}$ with respect to the basis coefficients is given by:
\begin{equation*}
\begin{split}
    \nabla\mathcal{L}_C =  \frac{\partial \mathcal{L}}{\partial c_{kl}} & = \frac{1}{n} \sum_{i=1}^{n} \left[\frac{\partial  \tr(\hat{\mathbf \Sigma}^{-1}_{i} \mathbf S_i)}{\partial c_{kl}} + \frac{\partial \log|\hat{\mathbf \Sigma}_{i}|}{\partial c_{kl}}\right],\\
       & = \frac{1}{n} \sum_{i=1}^{n} \left[ \tr\left(-\hat{\mathbf \Sigma}^{-1}_{i}\frac{\partial \hat{\mathbf \Sigma}_{i}}{\partial c_{kl}} \mathbf S_i \hat{\mathbf \Sigma}^{-1}_{i}\right) +  \tr\left(\hat{\mathbf \Sigma}^{-1}_{i}\frac{\hat{\partial \mathbf \Sigma}_{i}}{\partial c_{kl}}\right)\right],
\end{split}
\end{equation*}
where $\frac{\partial \hat{\mathbf \Sigma}_{i}}{\partial c_{kl}}$ is given as:

\begin{equation*}
\begin{split}
   \frac{\partial \hat{\mathbf \Sigma}_{i}}{\partial c_{kl}} &= \frac{\partial \boldsymbol{\Phi_i}\boldsymbol{\Lambda}\boldsymbol{\Phi_i}^\top + \mathbf I\sigma^2}{\partial c_{kl}}, \\
   & = \frac{\partial \boldsymbol{\Phi_i}}{\partial c_{kl}} \boldsymbol{\Lambda}\boldsymbol{\Phi_i} + \boldsymbol{\Phi_i}\boldsymbol{\Lambda} \frac{\partial \boldsymbol{\Phi_i}^\top}{\partial c_{kl}},\\
   & = \left(\frac{\partial \boldsymbol{\Phi_i}}{\partial \boldsymbol{u}_k} \times \frac{\partial \boldsymbol{u}_k}{\partial c_{kl}}\right) \boldsymbol{\Lambda}\boldsymbol{\Phi_i} + \boldsymbol{\Phi_i}\boldsymbol{\Lambda} \left(\frac{\partial \boldsymbol{\Phi_i}}{\partial \boldsymbol{u}_k} \times \frac{\partial \boldsymbol{u}_k}{\partial c_{kl}}\right)^\top,\\
   & = \left(\frac{\partial \boldsymbol{\Phi_i}}{\partial \boldsymbol{u}_k} \times \delta_{lk} \mathbf B_k\right) \boldsymbol{\Lambda}\boldsymbol{\Phi_i} + \boldsymbol{\Phi_i}\boldsymbol{\Lambda} \left(\frac{\partial \boldsymbol{\Phi_i}}{\partial \boldsymbol{u}_k} \times \delta_{lk} \mathbf B_k\right)^\top.
\end{split}
\end{equation*}

where \(\frac{\partial \phi_k}{\partial \boldsymbol{u}_v}\) is given as:
\begin{equation}\label{DWDU}
    \frac{\partial \phi_k}{\partial \boldsymbol{u}_v} = \left(\frac{\mathbf I_M}{\|\boldsymbol{w}_k\|_2} - \frac{\boldsymbol{w}_k\boldsymbol{w}_k^T}{\|\boldsymbol{w}_k\|^3_2}\right) \frac{\partial \boldsymbol{w}_k}{\partial \boldsymbol{u}_v},
\end{equation}
where \(\boldsymbol{w}_k = \boldsymbol{u}_k - \sum_{v = 1}^{k-1}\left(\frac{\boldsymbol{w}_v^T\boldsymbol{u}_k}{\boldsymbol{w}_v^T\boldsymbol{w}_v}\right)\boldsymbol{w}_v\) with \(k = 1,\cdots, p\) are orthogonal basis of \(\{\boldsymbol{u}_1,\ldots, \boldsymbol{u}_p\}\).
The derivative in (\ref{DWDU}) was derived in \cite{Edeling2021}. 
\subsection*{A.2.2. Lambda}
The derivative of the average negative log-likelihood loss $\mathcal{L}$ with respect to the eigenvalues is given by:
\begin{equation*}
\begin{split}
    \nabla\mathcal{L}_{\lambda_k} = \frac{\partial \mathcal L}{\partial \lambda_{k}} & = \frac{1}{n} \sum_{i=1}^{n} \left[\frac{\partial  \tr(\hat{\mathbf \Sigma}^{-1}_{i} \mathbf S_i)}{\partial \lambda_{k}} + \frac{\partial \log|\hat{\mathbf \Sigma}_{i}|}{\partial \lambda_{k}}\right],\\
     & = \frac{1}{n} \sum_{i=1}^{n} \left[ \tr\left(\frac{\partial \hat{\mathbf \Sigma}^{-1}_{i} \mathbf S_i}{\partial \lambda_{k}}\right)  +  \tr\left(\hat{\mathbf \Sigma}^{-1}_{i}\frac{\hat{\partial \mathbf \Sigma}_{i}}{\partial \lambda_{k}}\right)\right],\\
       & = \frac{1}{n} \sum_{i=1}^{n} \left[ \tr\left(-\hat{\mathbf \Sigma}^{-1}_{i}\frac{\partial \hat{\mathbf \Sigma}_{i}}{\partial \lambda_{k}} \mathbf S_i \hat{\mathbf \Sigma}^{-1}_{i}\right) +  \tr\left(\hat{\mathbf \Sigma}^{-1}_{i}\frac{\partial \hat{\mathbf \Sigma}_{i}}{\partial \lambda_{k}}\right)\right],
\end{split}
\end{equation*}
where $\frac{\partial \hat{\mathbf \Sigma}_{i}}{\partial \lambda_{k}}$ is given as: 
\begin{equation*}
\begin{split}
   \frac{\partial \hat{\mathbf \Sigma}_{i}}{\partial \lambda_{k}} &= \frac{\partial\boldsymbol{\Phi_i}\Lambda\boldsymbol{\Phi_i}^T + \mathbf I\sigma^2}{\partial \lambda_{k}} = \boldsymbol{\Phi}_{i} \frac{\partial \mathbf \Lambda}{\partial \lambda_k} \boldsymbol{\Phi}_{i}^T\\
\end{split}
\end{equation*}

where the partial derivative of $\boldsymbol{\Lambda}$ with respect to $\lambda_k$ is the $p \times p$ diagonal matrix
\[
\frac{\partial \boldsymbol{\Lambda}}{\partial \lambda_k}
=
\mathbf{E}_{kk},
\]
where $\mathbf{E}_{kk}$ denotes the matrix with a one in the $(k,k)$ entry and zeros elsewhere.

\subsection*{A.2.3. Error variance}
The derivative of the average negative log-likelihood loss $\mathcal{L}$ with respect to the error variance is given by:
\begin{equation*}
\begin{split}
    \frac{\partial \mathcal{L}}{\partial \sigma^2} & = \frac{1}{n} \sum_{i=1}^{n} \left[ \frac{\partial  \tr(\hat{\mathbf \Sigma}^{-1}_{i} \mathbf S_i)}{\partial \sigma^2} + \frac{\partial \log|\hat{\mathbf \Sigma}_{i}|}{\partial \sigma^2} \right],\\
       & = \frac{1}{n} \sum_{i=1}^{n} \left[ \tr\left(-\hat{\mathbf \Sigma}^{-1}_{i}\frac{\partial \hat{\mathbf \Sigma}_{i}}{\partial \sigma^2} \mathbf S_i \hat{\mathbf \Sigma}^{-1}_{i}\right) +  \tr\left(\hat{\mathbf \Sigma}^{-1}_{i}\frac{\hat{\partial \mathbf \Sigma}_{i}}{\partial \sigma^2}\right) \right],\\
       & = \frac{1}{n} \sum_{i=1}^{n} \left[ \textrm{tr}\left(-\hat{\mathbf \Sigma}^{-1}_{i}\frac{\partial \sigma^2 \mathbf I_{m_i}}{\partial \sigma^2} \mathbf S_i \hat{\mathbf \Sigma}^{-1}_{i}\right) +  \tr\left(\hat{\mathbf \Sigma}^{-1}_{i}\frac{\partial \sigma^2 \mathbf I_{m_i}}{\partial \sigma^2}\right)\right],\\
       & = \frac{1}{n} \sum_{i=1}^{n} \left[ \textrm{tr}\left(-\hat{\mathbf \Sigma}^{-1}_{i} \mathbf S_i \hat{\mathbf \Sigma}^{-1}_{i}\right) +  \tr\left(\hat{\mathbf \Sigma}^{-1}_{i}\right)\right].
\end{split}
\end{equation*}

\section*{A.3. Convergence of the Proposed Estimator}

Let
$\mathbf W = \mathrm{diag}(w_1, \ldots, w_M)$ denote the $M \times M$ diagonal matrix of quadrature weights, $\mathbf B \in \mathbb{R}^{M \times Q}$ the fixed matrix of basis functions evaluated on the quadrature grid $\boldsymbol{\tau}$, and $\mathbf C = [c_1, \ldots, c_p] \in \mathbb{R}^{Q\times p}$ the coefficient matrix. Define the unconstrained basis expansion $\mathbf U = \mathbf{BC} \in \mathbb{R}^{M \times p}$. The MGS recursion in (6) orthonormalizes the columns of $\mathbf U$ under the discrete weighted inner product $\langle f, g \rangle_w = f^\top \mathbf W g$, yielding
$\boldsymbol{\Phi}(\textbf{C}) = \mathcal{M}_W(\mathbf{BC}) \in \mathbb{R}^{M \times p}$ satisfying $\mathbf{\Phi^\top W \Phi = I_p}$, where $\mathcal{M}_W(\cdot)$ denotes the orthogonal matrix obtained from the $\mathbf W$-weighted MGS map.

\begin{assumption}[Full Rank]
\label{ass:fullrank}
The matrix $\mathbf U = \mathbf B\mathbf C$ has full column rank $p$ for all $\mathbf C$ in the domain of optimisation, that is $\det(\mathbf U^\top \mathbf W \mathbf U) > 0$. 
\end{assumption}

\begin{assumption}[Distinct Eigenvalues]
  \label{ass:distinct}
  The eigenvalues satisfy $\lambda_1 > \lambda_2 > \cdots > \lambda_p > 0$.
\end{assumption}

\begin{assumption}[Bounded Eigenvalues]
\label{ass:eigenbound}
The eigenvalues satisfy $\lambda_k \geq \delta > 0$ for all 
$k = 1,\ldots,p$ and some fixed $\delta > 0$.
\end{assumption}

\begin{assumption}[Positive Parameters]
  \label{ass:positive}
  The error variance satisfies $\sigma^2 > 0$ and all eigenvalues satisfy $\lambda_k > 0$.
  This is enforced through the log-transformations $\gamma = \log\sigma^2$ and $\eta_k =
  \log\lambda_k$ introduced in Section~2.
\end{assumption}

\begin{assumption}[Sample Size]
  \label{ass:samplesize}
  The total number of observations satisfies $\sum_{i=1}^n m_i > Qp + p + 1$, combined with Assumption 1 ensures that the observed Fisher information is positive definite at the true parameter.
\end{assumption}

\begin{assumption}[Bounded Level Set]
  \label{ass:levelset}
  The sublevel set
  \begin{equation*}
    \mathcal{S}_0
    = \bigl\{(\mathbf \Phi, \mathbf \Lambda, \sigma^2) :
      \mathcal{L}(\mathbf \Phi, \mathbf \Lambda, \sigma^2)
    \leq \mathcal{L}(\mathbf \Phi_0, \mathbf \Lambda_0, \sigma_0^2)\bigr\}
  \end{equation*}
  is compact in the space of $(\mathbf \Phi(\textbf{C}), \mathbf \Lambda, \sigma^2)$.
\end{assumption}

The reduced-rank covariance at the quadrature grid is
\begin{equation*}
  \boldsymbol{\Sigma}
  = \mathbf \Phi(\textbf{C})\,\mathbf \Lambda\,\mathbf \Phi(\textbf{C})^\top + \sigma^2 \mathbf I_M,
  \qquad
  \Phi(\mathbf C) = \mathcal{M}_W(\mathbf{BC}).
\end{equation*}

\begin{proposition}[Identifiability]
  \label{prop:identifiability}
  Under Assumptions~\ref{ass:distinct}--\ref{ass:positive}, and assuming additionally that
  $\lambda_k \neq \sigma^2$ for all $k$, the parameters $(\mathbf \Lambda, \sigma^2)$ and the column space of $\mathbf \Phi(\textbf{C})$ are uniquely determined by $\mathbf \Sigma$. The matrix $\mathbf \Phi(\textbf{C})$
  itself is identified up to column sign flips.
\end{proposition}

\begin{proof}
  For a real symmetric matrix, eigenvectors corresponding to distinct eigenvalues are orthogonal, and if the eigenvalues are distinct and ordered, then each column $\boldsymbol{\Phi}_j$ of $\boldsymbol{\Phi}$ is uniquely determined up to sign \citep{lange1999numerical}. The $p$ eigenvalues of $\mathbf \Sigma$ strictly greater
  than $\sigma^2$ are exactly $\lambda_k + \sigma^2$ for $k = 1, \ldots, p$, and the remaining $M - p$ eigenvalues are equal to $\sigma^2$. Since $\lambda_1 > \cdots > \lambda_p >
  0$ and $\lambda_k \neq \sigma^2$. It follows that $\mathbf{\Lambda}$ and $\sigma^2$ are uniquely determined by $\mathbf{\Sigma}.$ Furthermore, since eigenvectors corresponding to distinct eigenvalues are unique up to sign, the columns of $\mathbf{\Phi}(\mathbf{C})$ are identified up to sign flips.
\end{proof}

\begin{remark}[Non-injectivity of the parametrisation]
  \label{rem:noninjectivity}
  The mapping from the coefficient matrix $\textbf{C}$ to the parameter $\boldsymbol{\Phi}$ of interest, defined as $\boldsymbol{\Phi}(\textbf{C}) = \mathcal M_W(\textbf{BC})$, is not injective. However, this does not violate identifiability of $(\mathbf \Phi, \mathbf \Lambda, \sigma^2)$. This non-injectivity stems from the fact that the column space of a matrix is invariant under right-multiplication by any upper triangular matrix $\textbf{R}$ with positive diagonal entries, we have $\mathcal M_W(\textbf{BCR}) = \mathcal M_W(\textbf{BC})$, and thus $\boldsymbol{\Phi}(\textbf{CR}) = \boldsymbol{\Phi}(\textbf{C})$. However, the likelihood $\mathcal{L}(\mathbf C, \mathbf \Lambda, \sigma^2)$
  depends on $\mathbf C$ only through $\mathbf \Phi(\textbf{C})$, so this redundancy does not affect the estimator of $(\mathbf \Phi, \mathbf \Lambda, \sigma^2)$, it only affects the representation of the solution in the space of $\textbf{C}$. Furthermore, the gradient of $\mathcal{L}$
  with respect to $\mathbf C$ is identically zero in the flat directions $\mathbf C \mapsto \mathbf C\mathbf R$.

\end{remark}

\begin{lemma}[MGS is Smooth]\label{lem:mgs_smooth}
  The weighted MGS map $\mathcal{M}_W : \mathbb{R}^{M \times p} \to
  \mathbb{R}^{M \times p}$ is infinitely differentiable on the open set
  \begin{equation*}
    \mathcal{U} := \{\mathbf U \in \mathbb{R}^{M \times p} :
    \det(\mathbf U^\top \mathbf W \mathbf U) > 0\}.
  \end{equation*}
\end{lemma}

\begin{proof}
  The MGS procedure proceeds inductively over $k = 1, \ldots, p$. At
  each step it applies two operations: a projection:
      $\tilde{\varphi}_k = u_k - \sum_{\nu=1}^{k-1}
      \langle u_k, \varphi_\nu \rangle_w \varphi_\nu$,
      which is a linear and hence smooth operation in $\mathbf U$ and a normalisation:
      $\varphi_k = \tilde{\varphi}_k /
      \|\tilde{\varphi}_k\|_w$,
      which is smooth wherever $\|\tilde{\varphi}_k\|_w > 0$.
  It remains to verify that $\|\tilde{\varphi}_k\|_w > 0$ throughout
  $\mathcal{U}$. The condition $\det(\mathbf U^\top \mathbf W \mathbf U) > 0$ implies that the
 columns of $\textbf{U}$ are linearly independent under $\langle\cdot,
  \cdot\rangle_w$. Since $\varphi_1, \ldots, \varphi_{k-1}$ span
  the same subspace as $u_1, \ldots, u_{k-1}$, linear independence
  implies $u_k \notin \mathrm{span}(\varphi_1, \ldots,
  \varphi_{k-1})$, so $\tilde\varphi_k \neq 0$ and
  $\|\tilde\varphi_k\|_w > 0$. As each step is a composition of
  smooth operations with non-vanishing denominators, the entire map
  $\mathcal{M}_W$ is smooth on $\mathcal{U}$ by induction and the
  chain rule. See, for example, \citep{Dieci1999}.
\end{proof}

\begin{lemma}[MGS is locally Lipschitz]\label{lem:mgs_lipschitz}
  On any compact and convex subset $K \subset \mathcal{U}$, the MGS map
  satisfies
  \begin{equation*}
    \|\mathcal{M}_W(\mathbf U_1) - \mathcal{M}_W(\mathbf U_2)\|_F
    \leq L_K \|\mathbf U_1 - \mathbf U_2\|_F
  \end{equation*}
  for some constant $L_K < \infty$ depending only on $K$.
\end{lemma}

\begin{proof}
  Since $\mathcal{M}_W$ is smooth on $\mathcal{U}$
  (Lemma~\ref{lem:mgs_smooth}), its derivative $D\mathcal{M}_W(\mathbf U)$
  is continuous on $\mathcal{U}$. On the compact set $K \subset \mathcal{U}$, continuity implies
  \begin{equation*}
    \sup_{U \in K} \|D\mathcal{M}_W(\mathbf U)\|_{\mathrm{op}}
    =: L_K < \infty,
  \end{equation*}
  where $\| \cdot \|_{\mathrm{op}}$ denotes the operator norm induced by the Frobenius norm. For any $\mathbf U_1, \mathbf U_2 \in K$, convexity ensures the line segment
  $\mathbf U_1 + t(\mathbf U_2-\mathbf U_1)$ stays in $K$ for all $t \in [0,1]$. Applying
  the integral form of the mean value theorem
  (see \cite{lang2012real} for example), taking Frobenius norms and using the operator norm bound produces:
  \begin{align*}
    \|\mathcal{M}_W(\mathbf U_1) - \mathcal{M}_W(\mathbf U_2)\|_F
    &\leq \int_0^1 \|D\mathcal{M}_W(\mathbf U_1 + t(\mathbf U_2 - \mathbf U_1))
    (\mathbf U_2 - \mathbf U_1)\|_F \, dt \\
    &\leq \sup_{t \in [0,1]}
    \|D\mathcal{M}_W(\mathbf U_1 + t(\mathbf U_2-\mathbf U_1))\|_{\mathrm{op}}
    \cdot \|\mathbf U_2 - \mathbf U_1\|_F \\
    &\leq L_K \|\mathbf U_2 - \mathbf U_1\|_F. \qedhere
  \end{align*}
  To prove local Lipschitz continuity on \(\mathcal U\), fix \(\mathbf U_0 \in \mathcal U\). Since \(\mathcal U\) is open, there exists \(r > 0\) such that the closed ball \(\overline{B}(\mathbf U_0, r) \subset \mathcal U\). Because \(\overline{B}(\mathbf U_0, r)\) is compact and convex, the preceding argument applies with \(K = \overline{B}(\mathbf U_0, r)\). Therefore, \(\mathcal M_W\) is locally Lipschitz at \(\mathbf U_0\), and since \(\mathbf U_0\) was arbitrary, it is locally Lipschitz on \(\mathcal U\).
\end{proof}

\begin{corollary}[Smoothness of the Objective]
  \label{cor:smoothness}
  Under Assumption~\ref{ass:fullrank}, the map $\textbf{C} \mapsto
  \mathcal{L}(\textbf{C}, \eta, \gamma)$ is continuously differentiable on its domain. Moreover, its gradient \(\nabla_{\mathbf C}\mathcal L(\mathbf C,\eta,\gamma)\) is Lipschitz in all compact convex subset of the domain.
\end{corollary}

\begin{proof}
  The full map is the composition
  $\mathbf C
    \xrightarrow{\;\text{linear}\;} \mathbf B\mathbf C
    \xrightarrow{\;\mathcal{M}_W\;} \mathbf \Phi(\textbf{C})
    \xrightarrow{\;\text{smooth}\;} \mathcal{L}.$ The first map $\mathbf C \mapsto \mathbf B\mathbf C$ is linear and hence smooth. Under Assumption~\ref{ass:fullrank}, $\mathbf B\mathbf C \in \mathcal{U}$ throughout  the optimisation, so the second map $\mathcal{M}_W$ is smooth by  Lemma~\ref{lem:mgs_smooth}. The third map $\mathbf \Phi \mapsto  \mathcal{L}$ is smooth whenever $\mathbf \Sigma_i$ is positive definite,  which holds under Assumptions~\ref{ass:distinct} and  \ref{ass:positive}. By the chain rule for $C^\infty$ functions, the composition $\textbf{C} \mapsto \mathcal{L}(\textbf{C}, \eta, \gamma)$ is $C^\infty$ on its domain. In particular, it is twice continuously differentiable, so its gradient $\nabla_\textbf{C}\mathcal{L}$ is $C^1$.

 To prove that the gradient is locally Lipschitz continuous, fix \(\mathbf C_0\) in the domain. For any $\mathbf C_1, \mathbf C_2$ in the closed ball $\overline{B}(\mathbf C_0, r)$, the mean value inequality for vector-valued functions gives:
\[
\|\nabla \mathcal{L}(\mathbf C_1) - \nabla \mathcal{L}(\mathbf C_2)\|_F \leq \sup_{\mathbf C \in [\mathbf C_1, \mathbf C_2]} \|D(\nabla \mathcal{L})(\mathbf C)\|_{\text{op}} \cdot \|\mathbf C_1 - \mathbf C_2\|_F.
\]
The derivative $D(\nabla \mathcal{L})$ is continuous on the compact set $\overline{B}(\mathbf C_0, r)$, hence bounded. Let $L = \sup_{\mathbf C \in \overline{B}(\mathbf C_0, r)} \|D(\nabla \mathcal{L})(\mathbf C)\|_{\text{op}}$. Then $\|\nabla \mathcal{L}(\mathbf C_1) - \nabla \mathcal{L}(\mathbf C_2)\|_F \leq L \|\mathbf C_1 - \mathbf C_2\|_F$, proving Lipschitz continuity on the ball \citep{ferrera2013introduction}.
\end{proof}

\begin{proposition}[Compactness of Level Sets]
  \label{prop:compactness}
  Under Assumptions~\ref{ass:fullrank}--\ref{ass:samplesize}, the sublevel set
  $\mathcal{S}_0$ defined in Assumption~\ref{ass:levelset} is compact, thereby verifying
  Assumption~\ref{ass:levelset}.
\end{proposition}

\begin{proof}
  We show that $\mathcal{L} \to +\infty$ whenever the parameters approach the boundary of the parameter space, establishing that $\mathcal{S}_0$ is bounded. Closedness follows from the continuity of $\mathcal{L}$ under Assumption~\ref{ass:fullrank}. Recall the average
  negative log-likelihood:
  \begin{equation}\label{L}
    \mathcal{L}
    = \frac{1}{n}\sum_{i=1}^n
    \Bigl[
      \mathrm{tr}\bigl(\mathbf \Sigma_i^{-1}\textbf{S}_i)
      + \log|\mathbf \Sigma_i|
    \Bigr],
  \end{equation}
  where $\mathbf \Sigma_i = \mathbf \Phi_i \mathbf \Lambda \mathbf \Phi_i^\top + \sigma^2 \textbf{I}_{m_i}$ and $\textbf{S}_i$ is the sample covariance $\mathbf S_i =(\textbf{Y}_i - \boldsymbol{\mu}_i)(\textbf{Y}_i - \boldsymbol{\mu}_i)^\top$. The Woodbury matrix identity (see for example \citep{Golub2013}), gives 
  \begin{equation}\label{Wood}
    \mathbf \Sigma_i^{-1}
    = \frac{1}{\sigma^2}
    \Bigl(
      \textbf{I}_{m_i}
      - \mathbf \Phi_i\bigl(\mathbf \Lambda^{-1}\sigma^2 + \mathbf \Phi_i^\top\mathbf \Phi_i\bigr)^{-1}\mathbf \Phi_i^\top
    \Bigr)
  \end{equation}
  As $|\boldsymbol{\Sigma}_i| = \prod_{j=1}^p (\lambda_j + \sigma^2) \cdot \sigma^{2(m_i-p)}$,  the log‑determinant term  is
\begin{equation}\label{log_det}
\log | \boldsymbol{\Sigma}_i| = \sum_{j=1}^{p} \log(\lambda_j + \sigma^2) + (m_i - p)\log\sigma^2.
\end{equation}
  We consider all boundary cases in turn. 

  \medskip\noindent\textbf{Case~1: $\sigma^2 \to 0^+$.}
  As the noise variance shrinks to zero, $\mathbf \Sigma_i$ becomes nearly singular since
  $\lambda_{\min}(\mathbf \Sigma_i) \geq \sigma^2 \to 0$. Since $\boldsymbol{\Phi}_i^\top \boldsymbol{\Phi}_i = \textbf{I}_p$, as $\sigma^2 \to 0$ we have from (\ref{Wood}) $(\boldsymbol{\Lambda}^{-1}\sigma^2 + \textbf{I}_p)^{-1} \to \textbf{I}_p$, and therefore $(
      \textbf{I}_{m_i}
      - \mathbf \Phi_i\bigl(\mathbf \Lambda^{-1}\sigma^2 + \mathbf \Phi_i^\top\mathbf \Phi_i\bigr)^{-1} \boldsymbol{\Phi}_i^\top
    \Bigr) \rightarrow (
      \textbf{I}_{m_i}
      -\boldsymbol{\Phi}_i\boldsymbol{\Phi}_i^\top)=:\textbf{P}_i.$
Then, \[\tr{(\mathbf \Sigma_i^{-1}\mathbf S_i)} = (\mathbf Y_i - \mathbf \mu_i)^\top \mathbf \Sigma_i^{-1} (\mathbf Y_i - \mathbf \mu_i) \approx \frac{1}{\sigma^2} \|\mathbf P_i(\mathbf Y_i - \mathbf \mu_i)\|^2\]
This grows as $1/\sigma^2$ provided 
$\mathbf P_i(\mathbf Y_i - \mathbf \mu_i) \neq 0$ for some $i$. Since $\|\textbf{P}_i(\textbf{Y}_i - \boldsymbol{\mu}_i)\|^2 \geq 0$ for each $i$, the sum equals zero only if $\mathbf Y_i - \boldsymbol{\mu}_i \in \mathrm{span}(\boldsymbol{\Phi}_i)$ for all $i$ simultaneously, meaning the model perfectly fits all observations with zero residual. This requires $\sum_i m_i \leq Qp + p + 1$, which is ruled out by Assumption~\ref{ass:samplesize}.
  As $\sigma^2 \rightarrow 0,$ the log‑determinant term satisfies $\log |\boldsymbol{\Sigma}_i| = O(\log \sigma^2)$, which is negligible compared to $1/\sigma^2$. Hence $\mathcal{L} \to +\infty$.

  \medskip\noindent\textbf{Case~2: $\sigma^2 \to +\infty$.}
  As the noise variance grows without bound, $\mathbf \Sigma_i$ becomes very large. Since all eigenvalues of $\mathbf \Sigma_i$ are at least $\sigma^2$, we have $|\mathbf \Sigma_i| \geq
  (\sigma^2)^{m_i}$, and from (\ref{log_det}) $\log|\mathbf \Sigma_i| \to +\infty$. The quadratic term $\operatorname{tr}(\boldsymbol{\Sigma}_i^{-1} \textbf{S}_i)$ is bounded (it tends to $0$ as $\sigma^2\to\infty$), therefore $\mathcal{L} \to +\infty$.

  \medskip\noindent\textbf{Case~3:} $\lambda_k \to 0^+$ for some $k$.
 As $\lambda_k \to 0$ with $\sigma^2 > 0$ fixed, $\log(\lambda_k + \sigma^2) \to \log \sigma^2$, which is finite. Hence (\ref{log_det}) remains bounded. From (\ref{Wood}), as $\lambda_k \to 0$, the $k$-th entry tends to $0$, while the other entries remain constant. Consequently $\boldsymbol{\Sigma}_i^{-1}$ converges to a finite matrix, so $\operatorname{tr}(\boldsymbol{\Sigma}_i^{-1} \textbf{S}_i)$ is bounded. Hence $\mathcal{L}$ tends to a finite limit, so the sublevel 
set is not compact without further restriction. Under 
Assumption~\ref{ass:eigenbound}, $\lambda_k \geq \delta > 0$  for all $k$, so this boundary cannot be reached and the 
sublevel set remains compact.

  \medskip\noindent\textbf{Case~4:} $\lambda_k \to +\infty$ for some $k$.
  As $\lambda_k \to +\infty$, from (\ref{Wood}), the $k$-th entry tends to $1$, while the other entries remain constant ($\lambda_j/(\sigma^2 + \lambda_j)),$ in the $\phi_{ik}$-direction the inverse variance goes to $0,$ while the other entries remain constant. As $\lambda_k \to +\infty,$ from (\ref{log_det}) we have $\log|\mathbf \Sigma_i| \geq \log(\lambda_k + \sigma^2) \to +\infty,$ while $\operatorname{tr}(\boldsymbol{\Sigma}_i^{-1} \textbf{S}_i)$ remains bounded (as $\boldsymbol{\Sigma}_i^{-1}$ converges to a finite matrix). Hence $\mathcal{L} \to +\infty$. 

  \medskip\noindent\textbf{Case~5:} $\mathbf \Phi$ is bounded.
By construction, $\boldsymbol{\Phi}$ is obtained from the MGS map $\mathcal{M}_W$ applied to $\textbf{B}\textbf{C}$, and $\mathcal{M}_W$ always returns a matrix with orthonormal columns (with respect to $\langle\cdot,\cdot\rangle_W$). Thus $\boldsymbol{\Phi}$ lies on the Stiefel manifold $\operatorname{St}(M,p)$, which is compact (closed and bounded in $\mathbb{R}^{M\times p}$)
  \citep{Absil2008}. Therefore $\boldsymbol{\Phi}$ can never approach the boundary, it is automatically confined to a compact set regardless of how large $\textbf{C}$ becomes.

  \medskip
  Under Assumptions~\ref{ass:fullrank}--\ref{ass:samplesize}, the only ways parameters can approach the boundary are covered by Cases 1, 2, 4, and 5, all of which drive $\mathcal{L} \to +\infty$. Therefore any sublevel set ${\mathcal{L} \le \mathcal{L}_0}$ cannot contain a sequence tending to the boundary; it must be contained in a bounded region. Since $\mathcal{L}$ is continuous, this sublevel set is also closed, hence compact.
\end{proof}

\begin{remark}
  \label{rem:compactness}
  Proposition~\ref{prop:compactness} is stated in terms of $(\mathbf \Phi, \mathbf \Lambda, \sigma^2)$
rather than $(\mathbf C, \mathbf \Lambda, \sigma^2)$ because, as noted in
Remark~\ref{rem:noninjectivity}, the sublevel set is not bounded in $C$-space due to the non-injectivity of $\mathbf C \mapsto \mathbf \Phi(\textbf{C})$. Since the objective depends on $\mathbf C$ only through the compact manifold-valued $\mathbf \Phi(\textbf C)$, this presents no practical difficulty: BFGS
iterates in $\mathbf \Phi$-space remain in a compact set throughout the optimisation.
\end{remark}

\begin{proposition}[Positive Definite Fisher Information]
\label{prop:fisher-positive}
Under Assumptions~\ref{ass:fullrank}--\ref{ass:samplesize}, and assuming the true covariance has the form \(\boldsymbol{\Sigma}_0 = \boldsymbol{\Phi}_0\boldsymbol{\Lambda}_0\boldsymbol{\Phi}_0^\top + \sigma_0^2 \textbf{I}\) with \(\lambda_{0,1} > \cdots > \lambda_{0,p} > 0\) and \(\sigma_0^2 > 0\), the Fisher information matrix of the Gaussian model with respect to \(\boldsymbol{\theta}=(\operatorname{vec}(\boldsymbol{\Phi}), \operatorname{diag}(\boldsymbol{\Lambda}), \sigma^2)\) is positive definite at the true parameter.
\end{proposition}

\begin{proof}
Under Assumptions~\ref{ass:fullrank}--\ref{ass:samplesize}, the observations are
i.i.d.\ $\mathbf{y}_i \sim \mathcal{N}(\mathbf{0},\boldsymbol{\Sigma}(\boldsymbol{\theta}))$
with
\[
\boldsymbol{\Sigma}(\boldsymbol{\theta})
= \boldsymbol{\Phi}\boldsymbol{\Lambda}\boldsymbol{\Phi}^{\top} + \sigma^2\mathbf{I}_M,
\]
where $\boldsymbol{\Phi}\in\mathbb{R}^{M\times p}$ satisfies
$\boldsymbol{\Phi}^{\top}\mathbf{W}\boldsymbol{\Phi}=\mathbf{I}_p$,
$\boldsymbol{\Lambda}=\operatorname{diag}(\lambda_1,\dots,\lambda_p)$, and $\sigma^2>0$.
The true parameter is
$\boldsymbol{\theta}_0=(\operatorname{vec}(\boldsymbol{\Phi}_0),
\operatorname{diag}(\boldsymbol{\Lambda}_0),\sigma_0^2)$
with $\lambda_{0,1}>\cdots>\lambda_{0,p}>0$ and $\sigma_0^2>0$.

The Fisher information matrix for a Gaussian covariance model is 
\[
\mathcal{I}(\boldsymbol{\theta})
= \tfrac{1}{2}\,\mathbf{J}(\boldsymbol{\theta})^{\top}
  \bigl(\boldsymbol{\Sigma}^{-1}\otimes\boldsymbol{\Sigma}^{-1}\bigr)
  \mathbf{J}(\boldsymbol{\theta}),
\]
where $\mathbf{J}(\boldsymbol{\theta})=\partial\operatorname{vech}(\boldsymbol{\Sigma})/\partial\boldsymbol{\theta}^{\top}$ \citep{magnus2019matrix}.
Since $\boldsymbol{\Sigma}^{-1}\otimes\boldsymbol{\Sigma}^{-1}$ is positive definite,
$\mathcal{I}(\boldsymbol{\theta}_0)$ is positive definite if and only if
$\mathbf{J}(\boldsymbol{\theta}_0)$ has full column rank, i.e.\ the map
$\boldsymbol{\theta}\mapsto\boldsymbol{\Sigma}$ is locally injective at $\boldsymbol{\theta}_0$.
We establish this injectivity by showing that any tangent vector
$(\boldsymbol{\Delta},\,\mathrm{d}\boldsymbol{\Lambda},\,\mathrm{d}\sigma^2)$ with
$\mathrm{d}\boldsymbol{\Sigma}=\mathbf{0}$ must be zero.

The matrix $\boldsymbol{\Sigma}_0=\boldsymbol{\Phi}_0\boldsymbol{\Lambda}_0\boldsymbol{\Phi}_0^{\top}+\sigma_0^2\mathbf{I}_M$
has $p$ distinct eigenvalues $\mu_k=\lambda_{0,k}+\sigma_0^2$, $k=1,\dots,p$,
and a $(M{-}p)$-fold eigenvalue $\sigma_0^2$.
Let $\mathbf{E}=[\mathbf{E}_1,\mathbf{E}_2]\in\mathbb{R}^{M\times M}$ be an orthonormal
eigenbasis of $\boldsymbol{\Sigma}_0$, with $\mathbf{E}_1\in\mathbb{R}^{M\times p}$ spanning
the signal subspace and $\mathbf{E}_2\in\mathbb{R}^{M\times(M-p)}$ spanning the noise
subspace, so that $\mathbf{E}^{\top}\mathbf{E}=\mathbf{I}_M$.
Since $\boldsymbol{\Phi}_0$ and $\mathbf{E}_1$ span the same $p$-dimensional subspace,
there exists an invertible $\mathbf{P}\in\mathbb{R}^{p\times p}$ such that
$\boldsymbol{\Phi}_0=\mathbf{E}_1\mathbf{P}$, and the constraint
$\boldsymbol{\Phi}_0^{\top}\mathbf{W}\boldsymbol{\Phi}_0=\mathbf{I}_p$ gives
$\mathbf{P}^{\top}\mathbf{G}\mathbf{P}=\mathbf{I}_p$, where
$\mathbf{G}=\mathbf{E}_1^{\top}\mathbf{W}\mathbf{E}_1$ is positive definite.

Write the tangent direction as $\boldsymbol{\Delta}=\mathbf{E}_1\mathbf{A}+\mathbf{E}_2\mathbf{B}$
with $\mathbf{A}\in\mathbb{R}^{p\times p}$ and $\mathbf{B}\in\mathbb{R}^{(M-p)\times p}$.
The first-order change of $\boldsymbol{\Sigma}$ is
\begin{equation}\label{eq:dSig}
\mathrm{d}\boldsymbol{\Sigma}
= \boldsymbol{\Delta}\boldsymbol{\Lambda}_0\boldsymbol{\Phi}_0^{\top}
  + \boldsymbol{\Phi}_0\,\mathrm{d}\boldsymbol{\Lambda}\,\boldsymbol{\Phi}_0^{\top}
  + \boldsymbol{\Phi}_0\boldsymbol{\Lambda}_0\boldsymbol{\Delta}^{\top}
  + \mathrm{d}\sigma^2\,\mathbf{I}_M.
\end{equation}
Setting $\mathrm{d}\boldsymbol{\Sigma}=\mathbf{0}$ and projecting via
$\mathbf{E}^{\top}\boldsymbol{\Phi}_0=[\mathbf{P}^{\top},\mathbf{0}^{\top}]^{\top}$
yields three block equations:
\begin{eqnarray}\label{eq:blocks}
\nonumber
\mathbf{E}_2^{\top}\,\mathrm{d}\boldsymbol{\Sigma}\,\mathbf{E}_2
  &=& \mathrm{d}\sigma^2\,\mathbf{I}_{M-p} = \mathbf{0}
  \;\Rightarrow\; \mathrm{d}\sigma^2 = 0,\\[2pt]
\nonumber
\mathbf{E}_2^{\top}\,\mathrm{d}\boldsymbol{\Sigma}\,\mathbf{E}_1
  &=& \mathbf{B}\boldsymbol{\Lambda}_0\mathbf{P}^{\top} = \mathbf{0}
  \;\Rightarrow\; \mathbf{B} = \mathbf{0},\\[2pt]
\mathbf{E}_1^{\top}\,\mathrm{d}\boldsymbol{\Sigma}\,\mathbf{E}_1
  &=& \mathbf{A}\boldsymbol{\Lambda}_0\mathbf{P}^{\top}
      + \mathbf{P}\,\mathrm{d}\boldsymbol{\Lambda}\,\mathbf{P}^{\top}
      + \mathbf{P}\boldsymbol{\Lambda}_0\mathbf{A}^{\top} = \mathbf{0},
\end{eqnarray}
where $\mathbf{B}=\mathbf{0}$ follows because $\boldsymbol{\Lambda}_0$ and $\mathbf{P}$ are
invertible. Hence $\boldsymbol{\Delta}=\mathbf{E}_1\mathbf{A}$.

Substituting $\boldsymbol{\Delta}=\mathbf{E}_1\mathbf{A}$ into the linearised constraint
$\boldsymbol{\Phi}_0^{\top}\mathbf{W}\boldsymbol{\Delta}+\boldsymbol{\Delta}^{\top}\mathbf{W}\boldsymbol{\Phi}_0=\mathbf{0}$
and using $\boldsymbol{\Phi}_0=\mathbf{E}_1\mathbf{P}$ gives
$\mathbf{P}^{\top}\mathbf{G}\mathbf{A}+\mathbf{A}^{\top}\mathbf{G}\mathbf{P}=\mathbf{0}$.
Setting $\mathbf{X}=\mathbf{P}^{-1}\mathbf{A}$ and using $\mathbf{P}^{\top}\mathbf{G}\mathbf{P}=\mathbf{I}_p$
reduces this to $\mathbf{X}+\mathbf{X}^{\top}=\mathbf{0}$, i.e.\ $\mathbf{X}$ is
skew-symmetric. Substituting $\mathbf{A}=\mathbf{P}\mathbf{X}$ into (\ref{eq:blocks}),
multiplying on the left by $\mathbf{P}^{-1}$ and on the right by $\mathbf{P}^{-\top}$,
and using $\mathbf{X}^{\top}=-\mathbf{X}$ gives
\begin{equation}\label{eq:commutator}
\mathbf{X}\boldsymbol{\Lambda}_0 - \boldsymbol{\Lambda}_0\mathbf{X} + \mathrm{d}\boldsymbol{\Lambda} = \mathbf{0}.
\end{equation}
For $j\neq k$, the $(j,k)$ entry of (\ref{eq:commutator}) gives
$(\lambda_{0,k}-\lambda_{0,j})x_{jk}=0$.
Since the eigenvalues are distinct, $x_{jk}=0$ for all $j\neq k$, so $\mathbf{X}$ is
diagonal. A matrix that is both diagonal and skew-symmetric must be zero, hence
$\mathbf{X}=\mathbf{0}$. Equation~(\ref{eq:commutator}) then forces
$\mathrm{d}\boldsymbol{\Lambda}=\mathbf{0}$, and $\mathbf{A}=\mathbf{P}\mathbf{X}=\mathbf{0}$,
so $\boldsymbol{\Delta}=\mathbf{0}$.

Thus the only tangent vector yielding $\mathrm{d}\boldsymbol{\Sigma}=\mathbf{0}$ is the zero
vector, so the differential of $\boldsymbol{\theta}\mapsto\boldsymbol{\Sigma}$ is injective at
$\boldsymbol{\theta}_0$. Hence $\mathbf{J}(\boldsymbol{\theta}_0)$ has full column rank, and since
$\boldsymbol{\Sigma}_0^{-1}\otimes\boldsymbol{\Sigma}_0^{-1}$ is positive definite, the Fisher
information matrix $\mathcal{I}(\boldsymbol{\theta}_0)$ is positive definite.
\end{proof}

\begin{theorem}[Convergence of BFGS for Non-Convex Objectives]

Under Assumptions~\ref{ass:fullrank}--\ref{ass:levelset}: let $\{\boldsymbol{\theta}_k\}=\{(\textrm{vec}(\textbf{C}_k),\boldsymbol{\eta}_k \gamma_k)\},$ where $\gamma_k = \log \sigma_k^2$, ensuring positivity of the error variance and $\boldsymbol{\eta}_k = \log \boldsymbol{\lambda}_k,$ $\mathcal{L}(\boldsymbol{\theta})$ is twice continuously differentiable and bounded below, with Lipschitz continuous gradient on open convex sets containing the iterates. Let $\{\boldsymbol{\theta}_k\}$ be generated by the BFGS algorithm with a line search satisfying the Wolfe conditions (sufficient decrease and curvature). Assume the initial sublevel set $\mathcal{S}_0 = \{\boldsymbol{\theta} : \mathcal{L}(\boldsymbol{\theta}) \le \mathcal{L}(\boldsymbol{\theta}_0)\}$ is compact. Then:
\begin{enumerate}
    \item[(i)] \textbf{Convergence to stationarity:} Every limit point of $\{\boldsymbol{\theta}_k\}$ is a stationary point. If, in addition, the BFGS Hessian approximations remain uniformly positive definite with bounded condition number, then $\|\nabla \mathcal{L}(\boldsymbol{\theta}_k)\| \to 0$.
    \item[(ii)] \textbf{Local superlinear convergence:} If the sequence $\{\boldsymbol{\theta}_k\}$ converges to a point $\boldsymbol{\theta}^*$ that is a local minimizer with $\nabla^2 \mathcal{L}(\boldsymbol{\theta}^*)$ positive definite, then the rate of convergence is superlinear.
\end{enumerate}
\end{theorem}

\begin{proof}
\textbf{Part (i): Convergence to stationarity.} By the compactness of $\mathcal{S}_0$ and the descent property of the Wolfe line search, all iterates remain in $\mathcal{S}_0$. Zoutendijk's theorem \citep{nocedal2006numerical} applies under Lipschitz continuity of the gradient and the Wolfe conditions, yielding
\[
\sum_{k=0}^{\infty} \frac{(\nabla \mathcal{L}_k^\top \textbf{d}_k)^2}{\|\textbf{d}_k\|^2} < \infty. \tag{1}
\]
For BFGS, the search direction is $\textbf{d}_k = -\textbf{H}_k^{-1} \nabla \mathcal{L}_k$ with $\textbf{H}_k$ positive definite. If the matrices $\textbf{H}_k$ have uniformly bounded condition number, then the angle $\boldsymbol{\theta}_k$ between $-\nabla \mathcal{L}_k$ and $\textbf{d}_k$ satisfies $\cos \boldsymbol{\theta}_k \ge \delta > 0$ for all $k$ \citep[Lemma 4.1]{nocedal2006numerical}. Since $\cos \boldsymbol{\theta}_k = -(\nabla \mathcal{L}_k^\top \textbf{d}_k)/(\|\nabla \mathcal{L}_k\|\|\textbf{d}_k\|)$, we have $(\nabla \mathcal{L}_k^\top \textbf{d}_k)^2 / \|\textbf{d}_k\|^2 \ge \delta^2 \|\nabla \mathcal{L}_k\|^2$ \citep[Theorem 3.2]{nocedal2006numerical}. Substituting into (1) gives $\sum \|\nabla \mathcal{L}_k\|^2 < \infty$, hence $\|\nabla \mathcal{L}_k\| \to 0$. Without the bounded condition number, (1) only guarantees $\liminf \|\nabla \mathcal{L}_k\| = 0$; however, any limit point of $\{\boldsymbol{\theta}_k\}$ in the compact set $\mathcal{S}_0$ must then satisfy $\nabla \mathcal{L}(\boldsymbol{\theta}) = 0$ by continuity of the gradient. Thus every limit point is a stationary point.

\textbf{Part (ii): Local superlinear convergence.} Assume $\boldsymbol{\theta}_k \to \boldsymbol{\theta}^*$, where $\boldsymbol{\theta}^*$ is a local minimizer with $\nabla^2 \mathcal{L}(\boldsymbol{\theta}^*)$ positive definite. By continuity of the Hessian, there exists a neighborhood of $\boldsymbol{\theta}^*$ where $\nabla^2 \mathcal{L}$ is positive definite and $\mathcal{L}$ is strongly convex. For all sufficiently large $k$, the iterates lie in this neighborhood, and the Wolfe conditions are eventually satisfied with step length $\alpha_k = 1$ \citep[Theorem 6.6]{nocedal2006numerical}. For quasi-Newton methods with unit steps asymptotically, the Dennis–Moré theorem \citep{dennis1974characterization} states that convergence is superlinear if and only if
\[
\lim_{k\to\infty} \frac{\| (H_k - \nabla^2 \mathcal{L}(\boldsymbol{\theta}^*)) \textbf{d}_k \|}{\|\textbf{d}_k\|} = 0.
\]
BFGS satisfies this condition under the stated assumptions \citep[Theorem 6.6]{nocedal2006numerical}. Therefore, the convergence rate is superlinear.
\end{proof}

\begin{remark}
In non-convex optimization, one cannot guarantee convergence to a global or even a local minimizer. This is not specific to our proposed approach, the existing methods \citep{Peng2009} and \citep{He2022}, face the same theoretical limitation due to the non-convexity of $\mathcal{L}$ and the geometry of the Stiefel manifold \citep{Absil2008}. Part (i) establishes that any limit point of the BFGS iterates is stationary (gradient zero). This is the strongest global guarantee available without further assumptions. Part (ii) is a local result: if the sequence happens to attain a point where the Hessian is positive definite (a strict local minimiser), then the convergence from this point is superlinear. The additional condition of uniformly bounded condition number in Part (i) ensures that the search directions do not become orthogonal to the gradient, leading to full convergence of the gradient to zero; this condition may fail in practice if the iterates approach regions where the true Hessian is ill-conditioned. 
\end{remark}

\begin{remark}[R implementation]
\label{rem:rimplementation}
The convergence results in Theorem~1 are stated for BFGS with a line
search satisfying sufficient decrease and curvature conditions. In practice, the proposed method uses R's \texttt{optim} function with \texttt{method = "BFGS"}, which implements a cubic interpolation line search \citep{Nash1990} satisfying these conditions. The conclusions of Theorem~1 (i) therefore hold under the R implementation. The superlinear convergence result in part~(ii) relies on the Dennis--Mor\'{e} condition, which holds independently of the specific line search used, provided the iterates converge to a non-degenerate minimiser. Furthermore, the R implementation of BFGS is robust to the occasional violation of the positive definiteness condition that can occur in non-convex regions. Because $\mathcal{L}$ is locally convex around its minimizers, these conditions are sufficient to guarantee that the algorithm converges successfully from a wide range of starting values. The practical evidence for convergence of our method is strong it converged successfully in 100\% of simulation replicates across all parameter settings in Section~3. This empirical reliability, combined with the computational advantages of the MGS parametrisation, provides a practical justification for the proposed approach.
\end{remark}

\section*{A.4. Parameter Initialization, Convergence Criteria, and Numerical Stabilization}\label{app:Opp}

In our implementation, the basis coefficients $\mathbf{C}$ are initialized using independent draws from a uniform distribution on $[-1,1]$. The eigenvalues are initialized according to \(\lambda_k = \log\!\left( 100 - \frac{100 - 1}{p - 1}(k-1) \right)\), while the error variance parameter $\sigma^2$ is initialized to \(\log(0.01)\). Convergence of the quasi-Newton optimization is assessed using two complementary diagnostics. Specifically, convergence is declared when either (i) the infinity norm of the gradient falls below $10^{-6}$ or (ii) the absolute change in the average log-likelihood between successive iterations is less than $10^{-5}$. For numerical stability, each individual covariance matrix \(\boldsymbol{\Sigma}_i\) in (8) is stabilized using spectral regularization to ensure positive definiteness and well-conditioned matrix operations during the optimization. As the covariance structure admits a low-rank decomposition, the Woodbury matrix identity can be used to obtain the inverse in a computationally efficient and numerically stable manner.

\section*{A.5. Simulations}\label{Apendix_Sim}
\subsection*{A.5.1. Gaussian Process Parameters}
\begin{figure}[htp]
    \centering
    \includegraphics[width=\textwidth]{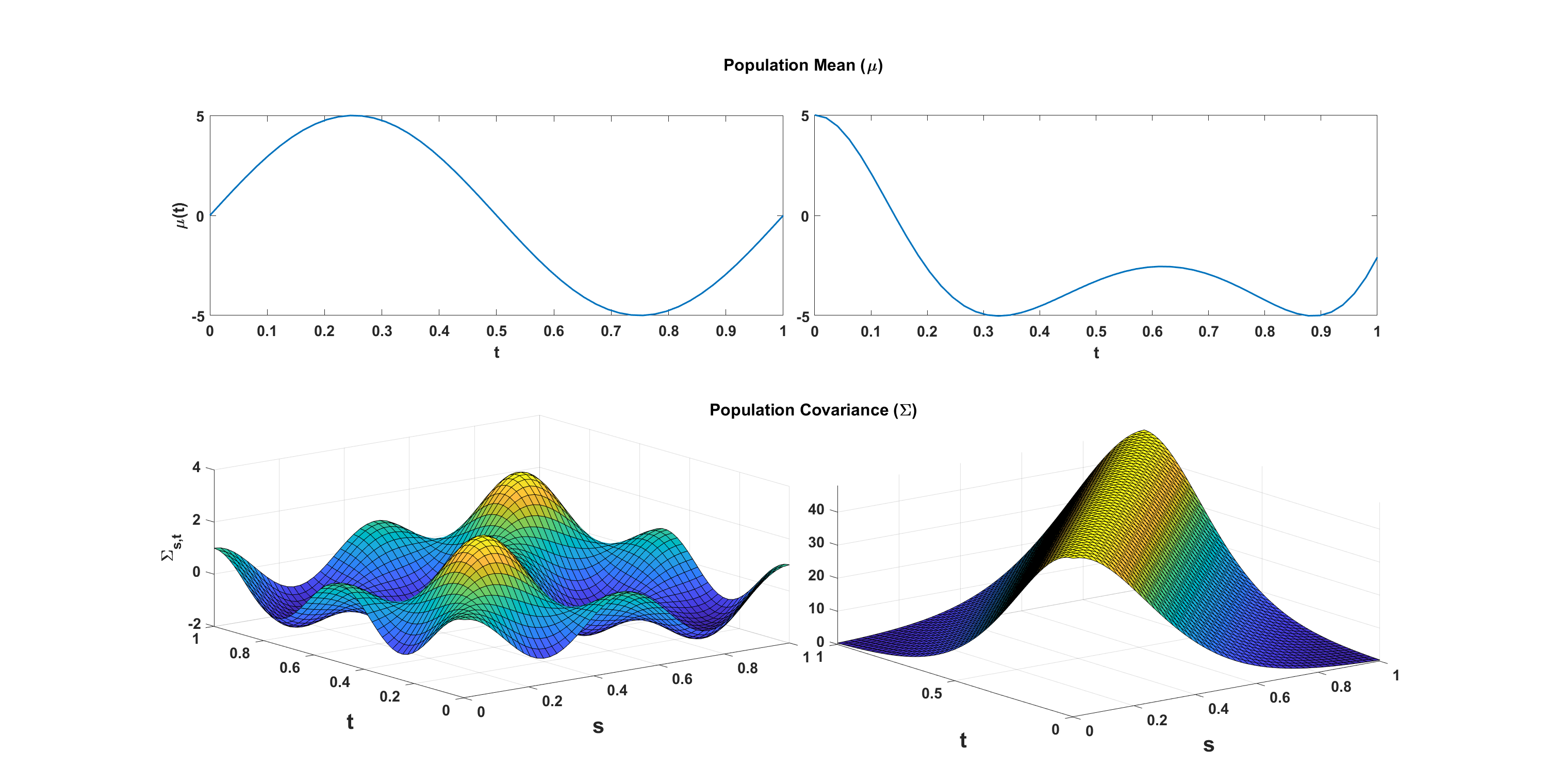}
    \caption{Population means and covariances for the stochastic processes used to generate the data are presented. The left panel shows the mean and covariance for the first distribution, referred to as the egg-crate example. The right panel displays the mean and covariance for the second distribution, known as the Matérn example. }
    \label{fig:sim_pop_pars}
\end{figure}

Figure \ref{fig:sim_pop_pars} presents the population means and covariances for the Gaussian processes used to generate the data in the simulation studies in Section 3 of the paper.

For the Matérn covariance scenarios, we used the mean function $\mu(t) = 5\cos(4t^3 + 6t^2 - 12t)$ (top right panel) and Matérn covariance parameters $\sigma = 1$, $\rho = 0.1$, and $\nu = 4$. For the finite basis expansion covariance (referred to as Egg-crate), we used the mean function $\mu(t) = 5\sin(2\pi t)$ (top left panel), with eigenfunctions $\phi_1(t) = \sqrt{2}\sin(2 \pi t)$, $\phi_2(t) = \sqrt{2}\cos(4 \pi t)$, $\phi_3(t) = \sqrt{2}\sin(4 \pi t)$, and eigenvalues $\lambda_1=1$, $\lambda_2=0.5$, and $\lambda_3=0.25$.
The Matérn Covariance (right bottom panel) is given by \(\Sigma(s,t) = \sigma^2 \frac{2^{1-\nu}}{\Gamma(\nu)}\left(\sqrt{2\nu} \frac{|s-t|}{\rho}\right)^{\nu} K_{\nu} \left(\sqrt{2\nu} \frac{|s-t|}{\rho}\right)\) and the finite basis expansion covariance (left bottom panel) \(\Sigma(s,t) = \sum^3_{k=1}\lambda_k \phi_k(s) \phi_k(t) \), where \(\phi_k\)'s are the eigenfunctions and \(\lambda_k\)'s are the eigenvalues.

\subsection*{A.5.2. Estimation Parameters for Simulations} \label{par}
The parameter specifications for the competing methods in the simulation studies described in Section 3 of the paper are as follows:
\begin{itemize}
\item Restricted Maximum Likelihood Estimate (ReMLE): For both the Matérn and egg-crate covariance surfaces, we used B-spline bases. We specified the number of possible basis functions and ranks to be the same as our proposed approach as described in the manuscript.
\item Conjugate Gradient (CG): We used the same specification provided on Github (\url{https://github.com/yehanxuan/CGPM}) and use the same number of possible basis functions and ranks as in our proposed approach.
\item Fast Covariance Estimation (FACE): We set the number of basis functions to 10 for both Matérn and egg-crate covariance surface simulations. The smoothing parameters were selected using generalized cross-validation (GCV) as recommended.
\item Principal Analysis by Conditional Expectations (PACE): We select the smoothing parameters using k-fold cross-validation as recommended.
\item Reproducing Kernel Hilbert Space (RKHS): The smoothing parameters were chosen using k-fold cross-validation as recommended.
\item Sparse Orthonormal Approximation (SOAP): Because smoothing parameter selections were unavailable, we instead used the penalty values provided in the SOAP GitHub implementation (\url{https://github.com/caojiguo/SOAP}
). These were 10 for the first eigenfunction and 30,000 for the second. Across all simulations, the method selected no more than two principal components. 
\end{itemize}

\subsection*{A.5.3 B-Spline Based Simulation}\label{app:Bs-spline_sim}
\begin{table}[htp!]
\centering
\caption{Median and  IQR of the Root Mean Squared Error of the top 3 Principal Eigenfunctions ($\text{RMSE}_{\phi_k}$) and Squared error of the top 3 Eigenvalues  ($\text{SE}_{\lambda_k}$), for 100 replicates of the B-Spline based simulations from the model selection section. The smallest median value in each column is highlighted in bold.
\label{table:rmse_PP_phi}} 
\resizebox{\textwidth}{!}{
 \begin{tabular}{lccccccc}
 \hline
 &\multicolumn{3}{c}{\textbf{n = 100}} && \multicolumn{3}{c}{\textbf{n = 500}}\\
\cline{2-4}
\cline{6-8}
  \multicolumn{8}{c}{\textbf{\(\text{RMSE}_{\phi_k}\)}} \\
 \hline
\textbf{Method}& \(\phi_1\) & \(\phi_2\) & \(\phi_3\) && \(\phi_1\) & \(\phi_2\) & \(\phi_3\) \\
\cline{2-4}
\cline{6-8}
\texttt{\(\text{mGSFPCA}_{Q\times p}\)} &  0.41 (0.38) & 0.76 (0.44) & 0.88 (0.50) && 0.13 (0.10) & 0.24 (0.24) & 0.21 (0.25) \\
\texttt{\(\text{mGSFPCA}_{Q + p}\)} &  0.38 (0.38) & 0.79 (0.48) & 0.90 (0.48) && 0.13 (0.10) & 0.24 (0.24) & 0.21 (0.24) \\
\texttt{ReMLE} &  0.39 (0.34) & \textbf{0.65 (0.44)} & 0.88 (0.38) && \textbf{0.12 (0.08)} & \textbf{0.21 (0.20)} & \textbf{0.18 (0.20)} \\
\texttt{CG} &  \textbf{0.36  (0.26)} & 0.67  (0.42) & \textbf{0.85  (0.54)} && 0.13  (0.08) & 0.23  (0.22) & 0.19  (0.23)\\ 
\texttt{FACE} &  1.04 (0.41) & 1.24 (0.27) & 1.16 (0.22) && 1.09 (0.46) & 1.16 (0.32) & 1.21 (0.27) \\
\texttt{PACE} &  0.96 (0.42) & 1.14 (0.33) & 1.16 (0.25) && 1.35 (0.11) & 1.36 (0.12) & 1.37 (0.07) \\
\texttt{RKHS} &  0.88 (0.41) & 1.18 (0.32) & 1.13 (0.24) && 0.57 (0.47) & 0.88 (0.52) & 0.93 (0.46) \\
  \hline
    \multicolumn{8}{c}{\textbf{\(\text{SE}_{\lambda_k} \times 100\)}} \\
 \hline
& \(\lambda_1\) & \(\lambda_2\) & \(\lambda_3\) && \(\lambda_1\) & \(\lambda_2\) & \(\lambda_3\) \\

\cline{2-4}
\cline{6-8}
\texttt{\(\text{mGSFPCA}_{Q\times p}\)} &  1.42 (4.38) & 0.74 (1.97) & 0.27 (0.63) && 0.22 (0.69) & \textbf{0.08 (0.21)} & 0.07 (0.26)\\ 
\texttt{\(\text{mGSFPCA}_{Q + p}\)} &  1.72 (6.48) & 0.87 (2.41) & 0.33 (0.66) && 0.22 (0.70) & \textbf{0.08 (0.22)} & 0.07 (0.26)\\ 
\texttt{ReMLE} &  \textbf{1.36 (4.41)} & 0.63 (1.71) & 0.18 (0.59) && \textbf{0.17 (0.58)} & \textbf{0.08 (0.22)} & \textbf{0.04 (0.20)}\\ 
\texttt{CG} &  1.47  (2.75) & \textbf{0.41  (0.93)} & \textbf{0.15  (0.64)} && 0.21  (0.51) & \textbf{0.08  (0.22)} & 0.09  (0.24)\\ 
\texttt{FACE} &  10.81 (10.17) & 5.90 (5.88) & 5.44 (3.91) && 12.58 (13.81) & 7.61 (7.25) & 8.10 (6.41)\\ 
\texttt{PACE} &  17.02 (20.8) & 13.47 (9.69) & 14.56 (5.95) && 23.27 (20.86) & 15.32 (10.77) & 16.31 (6.32)\\ 
\texttt{RKHS} &  4.75 (10.79) & 2.24 (5.99) & 2.55 (3.83) && 3.64 (8.10) & 3.21 (5.16) & 5.35 (5.56)\\ 
\hline
\end{tabular}   
}
\end{table}

Table \ref{table:rmse_PP_phi} summarizes the accuracy of the eigenvalues and eigenfunction estimation for 100 replicates of the B-Spline simulation, as described in Section 3.3 of the paper. The table presents the median and IQR of the Root Mean Squared Error (RMSE) between the true eigenfunctions and their estimated counterparts for the top three eigenfunctions \(\phi_1, \phi_2, \phi_3\), and \(100 \times\) Square Error (SE) between the true eigenvalues and estimated eigenvalues \(\lambda_1, \lambda_2, \lambda_3\).

The results suggest that \texttt{ReMLE}, \texttt{CG}, and \texttt{mGSFPCA} delivered competitive performance across most scenarios. Of these, \texttt{ReMLE} attained the lowest values in eight scenarios, with \texttt{CG} doing so in five. Nevertheless, for $n=500$, the differences among the methods were negligible. Our proposed method, \texttt{mGSFPCA}, maintained comparable accuracy, yielding values close to those of \texttt{ReMLE} and \texttt{CG}, while offering improved numerical stability. The robustness of \texttt{mGSFPCA} is particularly evident in its perfect convergence record: all 1200 models tested (100 replicates \(\times\) 12 possible models) converged successfully. In contrast, \texttt{ReMLE} exhibited convergence issues, with only 730 models converging for \(n = 100\) and 1108 for \(n = 500\) out of the same total. 

\newpage
\bibliographystyle{apalike}
\bibliography{ref.bib}